\documentclass[fleqn,usenatbib]{mnras}

\usepackage{newtxtext,newtxmath}
\usepackage{amsmath}

\usepackage[T1]{fontenc}
\usepackage{float}
\DeclareRobustCommand{\VAN}[3]{#2}
\let\VANthebibliography\thebibliography
\def\thebibliography{\DeclareRobustCommand{\VAN}[3]{##3}\VANthebibliography}

\usepackage{graphicx}	

\title[Lensing Effects on BSD during the EoR]{Gravitational Lensing Effects by Galaxy Clusters on Ionised Bubble Size Distribution during the Epoch of Reionisation}

\author[Wu et al.]{
Di Wu,$^{1,2}$
Nan Li,$^{1,2,3}$\thanks{E-mail: nan.li@nao.cas.cn (NL)}
Huanyuan Shan$^{4,5}$\thanks{E-mail: hyshan@shao.ac.cn (HS)}
and Zhenghao Zhu$^{4}$
\\
$^{1}$National Astronomical Observatories, Chinese Academy of Sciences, 20A Datun Road, Chaoyang District, Beijing 100101, China\\
$^{2}$School of Astronomy and Space Science, University of Chinese Academy of Sciences, Beijing 100049, China\\
$^{3}$Key Laboratory of Space Astronomy and Technology, National Astronomical Observatories, Chinese Academy of Sciences, 20A Datun Road, Chaoyang\\ District, Beijing 100101, China\\
$^{4}$Shanghai Astronomical Observatory, Chinese Academy of Sciences, 80 Nandan Road, Shanghai 200030, China\\
$^{5}$State Key Laboratory of Radio Astronomy and Technology, A20 Datun Road, Chaoyang District, Beijing, 100101, P. R. China
}

\date{Accepted XXX. Received YYY; in original form ZZZ}

\pubyear{2025}

\begin{document}
\label{firstpage}
\pagerange{\pageref{firstpage}--\pageref{lastpage}}
\maketitle

\begin{abstract}
The statistical properties of ionisation structures during the Epoch of Reionisation (EoR) provide valuable insights into the formation of the first stars and galaxies. However, size distributions of ionisation structures can be affected by gravitational lensing from foreground massive structures such as galaxy clusters. To quantify the impact of cluster lensing on the ionised Bubble Size Distribution (BSD), we performed a series of multiple-lens-plane simulations combining light cones of clusters with source light cones based on different ionisation models. The deflector population is generated with a Monte Carlo method guided by the halo mass function and empirical scaling relations, and the deflectors are modelled with truncated Navarro-Frenk-White (TNFW) profiles. Source light cones are produced semi-numerically or taken directly from the Evolution of 21 cm Structure (EOS) project. Using the Mean Free Path method, we measure both unlensed and lensed BSDs. We find that gravitational lensing increases the apparent abundance of large bubbles while leaving small bubbles nearly unchanged across all source models considered. In particular, for the EOS faint-galaxies model, the apparent number of bubbles with $R > 15\,\mathrm{cMpc}$ increases by 219\% at $z = 14$; for the EOS bright-galaxies model, it increases by 832\% under the same conditions. Moreover, a directional projection test shows only minor lensing-induced changes in line-of-sight direction, suggesting a possible route to recovering unlensed bubble statistics. Above all, lensing introduces unavoidable systematics into BSD measurements that should be carefully taken into account for relevant studies in the SKA era.

\end{abstract}




\section{Introduction}

The Epoch of Reionisation (EoR) is a pivotal period in the history of cosmic evolution, during which the universe underwent a global phase transition, shifting from a completely neutral state to one that is almost entirely ionised \citep{2005SSRv..116..625C, 2006PhR...433..181F, 2010ARA&A..48..127M, 2012RPPh...75h6901P, 2016ARA&A..54..761S, Dayal_2019, 2020PASP..132f2001L}. Observations and theoretical studies of the EoR are beneficial in understanding the properties of the first generation of stars \citep{2006ApJ...652....6Y, 2016ApJ...823..140X, 2023MNRAS.524..351K, 2024MNRAS.527.5102V} and galaxies \citep{2008PhRvD..78b3529M, 2008MNRAS.385L..63N, 2015Natur.519..327W, 2024MNRAS.535.1293J} in the Universe, comprehending the formation of large-scale cosmic structures \citep{Silva_2021}, and constraining cosmological models \citep{2006ApJ...653..815M, 2008PhRvD..78b3529M, castellano2023constraintsdarkmatterreionization, 2024MNRAS.528.2784D}.

Ionised bubbles are ionised regions in the intergalactic medium (IGM) during the EoR, which appear as dark areas in the 21 cm signal images of neutral hydrogen. The Bubble Size Distribution (BSD) represents the topology of reionisation \citep{2016MNRAS.461.3361L} and is one of the most important tracers for probing the progress of cosmic reionisation \citep{2004Natur.432..194W, 2020ApJ...891L..10T}. A large number of theoretical investigations have indicated that ionised bubbles form around ionising sources and grow in isolation in the initial stage, and as reionisation proceeds, different ionised bubbles merge, eventually leading to the complete ionisation of the IGM \citep{2007ApJ...654...12Z, 2007ApJ...669..663M, 2016MNRAS.457.1813F, 2022MNRAS.510.3858Q, 2024MNRAS.528.4872L}. Therefore, the evolution of BSD along redshift characterises the progress of reionisation, which is a powerful tool to constrain the parameters of reionisation models by comparing observations with theoretical predictions \citep{2007MNRAS.377.1043M, 2012MNRAS.420..441T, 2022A&A...667A.118D, 2024MNRAS.528.4872L}.

Among the tracers utilised to determine the properties of ionised bubbles during the EoR, the Lyman-$\alpha$ emission by high-redshift galaxies \citep{2006ApJ...647L..95M, 2014PASA...31...40D, 2018ApJ...856....2M} and the 21 cm signal in the radio band \citep{Datta_2007,2018MNRAS.473.2949G, Shimabukuro_2022,mishra2024detectingionizedbubblesluminous} are two notable representatives. The Lyman-$\alpha$ emission by high-redshift galaxies serves as an indirect tracer for inferring the morphology and size of ionised bubbles \citep{Endsley_2022, 2023MNRAS.524.5891T, Castellano_2016, Castellano_2018, Tilvi_2020, Larson_2022, tang2023jwstnirspecspectroscopyz79star}, as it is significantly attenuated by neutral hydrogen regions in the intergalactic medium (IGM), while ionised hydrogen regions do not absorb it. The direct tracer for assessing the morphology and size of ionised bubbles is the 21 cm emission from hydrogen itself in the radio band. Although the detection of the 21 cm emission from the EoR is challenging due to weak signals and considerable foreground contamination, plenty of efforts continue to advance this field, such as the Murchison Widefield Array (MWA) \footnote{MWA, \url{http://www.mwatelescope.org} } \citep{2013PASA...30....7T}, the 21 CentiMeter Array (21CMA) \footnote{21CMA, \url{https://english.nao.cas.cn/Research2015/rp2015/201701/t20170120_173603.html} } \citep{Zheng_2016}, the Low-Frequency Array (LOFAR) \footnote{LOFAR, \url{http://www.lofar.org} } \citep{2013A&A...556A...2V}, the Hydrogen Epoch of Reionization Array (HERA) \footnote{HERA, \url{https://reionization.org/} } \citep{DeBoer_2017}, and the Precision Array for Probing the Epoch of Reionization (PAPER) \footnote{PAPER, \url{http://eor.berkeley.edu} } \citep{Parsons_2010}. To date, the upper limit of the 21 cm signal power spectrum during the EoR has been explored extensively  \citep{ 2019ApJ...884....1B,2020MNRAS.493.1662M, 2021MNRAS.505.4775Y, 2022ApJ...925..221A, 2019ApJ...883..133K}, and the next-generation radio interferometer arrays like the Square Kilometre Array (SKA) are anticipated to directly observe the 21 cm signal from the EoR \citep{2015aska.confE..10M}.

Gravitational lensing distorts the morphology of background sources due to the deflection of light \citep{1992grle.book.....S, 1996astro.ph..6001N, 1998LRR.....1...12W, 2010CQGra..27w3001B, 2024SSRv..220...12S}. An effect which is known as magnification bias \citep{1980ApJ...242L.135T, 1991A&A...251L..35B, 1992grle.book.....S,2004MNRAS.351.1266W,von_Wietersheim_Kramsta_2021}, refers to the resulting selection bias caused by the lensing effect in flux- or size-limited samples, affecting the size or morphology distributions of high-redshift objects. For instance, the luminosity function of submillimeter galaxies, which are primarily distributed at redshifts 1 to 4, shows an observed number density at the bright end that is much higher than the theoretical predictions due to the lensing effects \citep{1996MNRAS.283.1340B, 2010ApJ...717L..31L, 2013MNRAS.430.1423E}. Similarly, the number density of bright UV galaxies at redshift around 8 has been significantly enhanced by gravitational lensing \citep{2015ApJ...805...79M}. Therefore, when utilising the BSD to infer the progress of reionisation and constrain the parameters of reionisation models, ignoring lensing effects may lead to systematic bias. Consequently, quantifying this impact is necessary, but it is still unclear.

In this paper, to quantify the gravitational lensing effects on BSD by clusters thoroughly, we conducted a series of multiple-lens-plane lensing simulations involving a light cone of clusters alongside source light cones based on various ionisation models. Expressly, the deflector population is generated using the Monte Carlo method, guided by the halo mass function \citep{2008ApJ...688..709T} which describes the halo density at different masses and redshifts, while the deflectors' mass profile is modelled using the Truncated Navarro-Frenk-White (TNFW) model \citep{1997ApJ...490..493N,2009JCAP...01..015B}. To explore the influences of lensing on BSD for different ionisation simulation models, we produce three types of source light cones via a semi-numerical approach by using Python package \texttt{21CMFAST v3} \citep{2016MNRAS.459.2342M, Murray2020}, at the same time two source light cones from the Evolution of 21 cm Structure (EOS) project \citep{2016MNRAS.459.2342M} are adopted. Subsequently, we run multi-lens-plane ray-tracing lensing simulations with the above mass and source light cones to generate two-dimensional (2D) \(x_{\mathrm{HI}}\) (neutral hydrogen fraction) maps with lensing effects. Following this, by comparing the BSDs, measured with the Mean Free Path (MFP) method \citep{2007ApJ...669..663M},  and their evolution along redshift from the 2D \(x_{\mathrm{HI}}\) maps with and without lensing effects, impacts of lensing on BSD can be quantified. Our findings reveal that the impacts of lensing are noteworthy on larger ionised bubbles at the early stage of EoR when redshift $\sim 12 - 14$ across all source models adopted, necessitating careful consideration in studies related to BSD in the era of the SKA.

The structure of this paper is as follows. Sect. \ref{section: theory} introduces the basics of reionisation and gravitational lensing relevant to this work. In Sect. \ref{section: simulation generation}, we detail the implementation of our simulations. Sect. \ref{section: Measuring BSD} describes the algorithm used to measure the BSD, and Sect. \ref{section: results} presents our findings. Lastly, we draw conclusions in Sect. \ref{section: conclusions}. Throughout this study, we adopt the $\Lambda$CDM cosmology, unless otherwise specified, the cosmological parameters are set based on the results from Planck 2018 \citep{2020A&A...641A...6P}: $\Omega_{\Lambda}=0.6889$,
$\Omega_{\rm m}=0.3111$, $\Omega_{\rm b}=0.04897$, $n_s=0.9665$, and $H_0 = 67.66$ km\,s$^{-1}$\,Mpc$^{-1}$.

\section{Theory}\label{section: theory}

In this section, we review the basic theory of the EoR and of gravitational lensing in Sect. \ref{section: EoR theory} and Sect. \ref{section: SL theory}, respectively.

\subsection{Basics of EoR simulation}\label{section: EoR theory}

The EoR simulations used in this work were generated with the publicly available Python software \texttt{21CMFAST} \citep{2011MNRAS.411..955M, 2016MNRAS.459.2342M, Murray2020}. \texttt{21CMFAST} employs a semi-numerical approach to model the evolution of key parameters during the EoR. In particular, the code simulates the 21 cm differential brightness temperature $\delta T_{b}(\nu)$ as given by \citet{2006PhR...433..181F}:
\begin{align}
\label{eq: delT}
\nonumber \delta T_{b}(\nu) \approx \nonumber&27x_{\rm HI} (1+\delta_{\rm nl}) \left(\frac{H(z)}{{\rm d}v_r/{\rm d} r + H(z)}\right) \left(1 - \frac{T_{\gamma}}{T_s} \right) \\
&\times \left( \frac{1+z}{10} \frac{0.15}{\Omega_{\rm m} h^2}\right)^{1/2} \left( \frac{\Omega_b h^2}{0.023} \right) {\rm mK},
\end{align}
where $T_s$ is the spin temperature of the neutral hydrogen gas, and $T_{\gamma}$ denotes the CMB photon temperature. We simulate the evolution of $T_s$ and $T_{\gamma}$ in this work instead of assuming the saturation of $T_s$. The quantity $\delta_{\rm nl}(\mathbf{x}, z) \equiv \rho/\bar{\rho} - 1$ is the density contrast; $H(z)$ is the Hubble parameter; ${\rm d}v_r/{\rm d} r$ is the comoving line-of-sight(LOS) velocity gradient; and $z$ is the redshift.

The $x_{\rm HI}$ is determined by the ionising influence of sources on their surrounding environment. This is modelled using an excursion-set formalism \citep{2004ApJ...613....1F, 2016MNRAS.459.2342M}, in which a cell is considered ionised if it satisfies the criterion:
\begin{equation}
\label{eq: ion_crit_coll}
\zeta f_{\rm coll}(\mathbf{x}, z, R, M_{\rm min}) \geq 1 + \bar{n}_{\rm rec}(\mathbf{x}, z, R).
\end{equation}
Here $f_{\rm coll}(\mathbf{x}, z, R, M_{\rm min})$ is the fraction of collapsed matter inside a spherical region of radius $R$ (centred at position $\mathbf{x}$), $\bar{n}_{\rm rec}(\mathbf{x}, z, R)$ is the cumulative number of recombinations \citep{2014MNRAS.440.1662S}, and $\zeta$ is the ionising efficiency (treated as a free parameter) \citep{2015MNRAS.449.4246G}. This excursion-set method effectively accounts for ionising photons from all sources hosted in haloes above the mass threshold $M_{\rm min}$.

The parameter $M_{\rm min}$ represents the minimum halo mass capable of hosting ionising sources. It is often defined via the halo’s virial temperature (i.e. the atomic cooling threshold). The relation between halo mass $M_{\rm halo}$ and virial temperature $T_{\rm vir}$ is given by \citet{2001PhR...349..125B}:
\begin{align}
M_{\rm halo} &= 10^{8} h^{-1} \left(\frac{w}{0.6}\right)^{-3/2}\left(\frac{\Omega_{\rm m}}{\Omega^{z}_{\rm m}}
\frac{\Delta{\rm c}}{18\pi^{2}}\right)^{-1/2} \nonumber \\
& \times \left(\frac{T_{\rm vir}}{1.98\times10^{4}~{\rm K}}\right)^{3/2}\left(\frac{1+z}{10}\right)^{-3/2}M_{\odot}.
\end{align}
Here $\Omega_{\mathrm{m}}^{z} \equiv \frac{\Omega_{\mathrm{m}}(1+z)^3}{\Omega_{\mathrm{m}}(1+z)^3+\Omega_{\Lambda}}$ is the effective matter density fraction at redshift $z$, $w$ is the mean molecular weight, and the collapse overdensity is $\Delta_{\rm c} = 18\pi^2 + 82d - 39d^2$ with $d \equiv \Omega_{\mathrm{m}}^{z} - 1$. Both $M_{\rm min}$ and $\zeta$ are crucial parameters for modelling the reionisation process. We discuss their impact on our simulations in Sect. \ref{section: simulation generation}.

\subsection{Basics of Gravitational Lensing}\label{section: SL theory}

We adopt the thin-lens approximation for the calculation of gravitational lensing \citep{1992grle.book.....S} in this paper, and the dimensionless projected surface mass density of the lens (the convergence) is given by:
\begin{equation}
\label{eq:kappa}
\kappa(\boldsymbol{\theta}) = \frac{\Sigma(\boldsymbol{\theta})}{\Sigma_{\mathrm{cr}}},
\end{equation}
where $\Sigma_{\mathrm{cr}}$ is the critical surface mass density given by:
\begin{equation}
\label{eq:sigma_cr}
\Sigma_{\mathrm{cr}} = \frac{c^2}{4\pi G} \frac{D_{\mathrm{s}}}{D_{\mathrm{d}}D_{\mathrm{ds}}}.
\end{equation}
Here $D_{\mathrm{s}}, D_{\mathrm{d}},$ and $D_{\mathrm{ds}}$ are the angular diameter distances from the observer to the source plane, from the observer to the lens plane, and from the lens plane to the source plane, respectively. $\Sigma(\boldsymbol{\theta})$ is the physical surface mass density of the lens, $c$ is the speed of light, and $G$ is the gravitational constant. The lensing potential $\psi(\boldsymbol{\theta})$ is defined by:
\begin{equation}
\label{eq:potential}
\psi(\boldsymbol{\theta}) = \frac{1}{\pi} \int_{\mathbb{R}^2} \kappa(\boldsymbol{\theta}') \ln\left|\boldsymbol{\theta} - \boldsymbol{\theta}'\right|d^2\theta',
\end{equation}
From this potential, one can derive the standard lensing quantities. For example, the (reduced) deflection angle is given by
\begin{equation}
\label{eq:deflection}
\boldsymbol{\alpha}(\boldsymbol{\theta}) = \nabla \psi(\boldsymbol{\theta}).
\end{equation}
The relationship between the source position $\boldsymbol{\beta}$ and the image position $\boldsymbol{\theta}$ is given by the lensing equation:
\begin{equation}
\label{eq:lensing_eq}
\boldsymbol{\beta} = \boldsymbol{\theta} - \boldsymbol{\alpha}(\boldsymbol{\theta}).
\end{equation}

This work employs the multiple-lens-plane ray-tracing approach. Specifically, $\kappa$ at position $\boldsymbol{\mathbf{\theta}}_i$ on the $i$th lens plane can be expressed by the following formula in  \citep{1992grle.book.....S}:
\begin{equation}
\label{eq:kappa_on_eachplane}
\kappa_i(\boldsymbol{\mathbf{\theta}}_i)=\frac{4\pi G}{c^2}\frac{D_iD_{is}}{D_s}\Sigma_i(D_i\mathbf{\boldsymbol{\theta}}_i),
\end{equation}
where $D_i$ and $D_{is}$ are the angular diameter distance from the observer to the $i$th lens plane and from the $i$th lens plane to the source plane, respectively. $\Sigma_i(D_i\boldsymbol{\mathbf{\theta}}_i)$ is the surface mass density at $D_i\boldsymbol{\mathbf{\theta}}_i$. Using Eq. \eqref{eq:potential} and Eq. \eqref{eq:deflection}, one can get the deflection angle $\boldsymbol{\mathbf{\alpha}}(\boldsymbol{\theta}_i)$ on the $i$th lens plane. Thus, the lensing equation Eq.~\eqref{eq:lensing_eq} becomes 
\begin{equation}
\boldsymbol{\mathbf{\theta}}_j=\boldsymbol{\mathbf{\theta}}_1-\sum_{i=1}^{j-1}\frac{D_{ij}D_{s}}{D_{j}D_{is}}\boldsymbol{\mathbf{\alpha}}(\boldsymbol{\theta}_i),
\end{equation}
where $\boldsymbol{\mathbf{\theta}}_j$ is the angular position of the light ray on the $j$th lens plane, $D_{j}$ and $D_{ij}$ are the angular diameter distance from observer to $j$th lens plane and from $i$th lens plane to $j$th lens plane.

\section{Simulations}\label{section: simulation generation}

This section comprehensively explains our simulations to generate lensed EoR maps to quantify the lensing effects on BSDs. Sect. \ref{section: Reionization simulation generation} details the generation of the reionisation simulations. Sect. \ref{section: Foreground deflector simulation} describes the construction of the light cones of deflectors. At last, Sect. \ref{section: lensing simulation} presents the implementation of the lensing simulations based on the above two components.

\subsection{EoR simulations}\label{section: Reionization simulation generation}

\begin{figure*}
\centering
\includegraphics[width=1.12\linewidth]{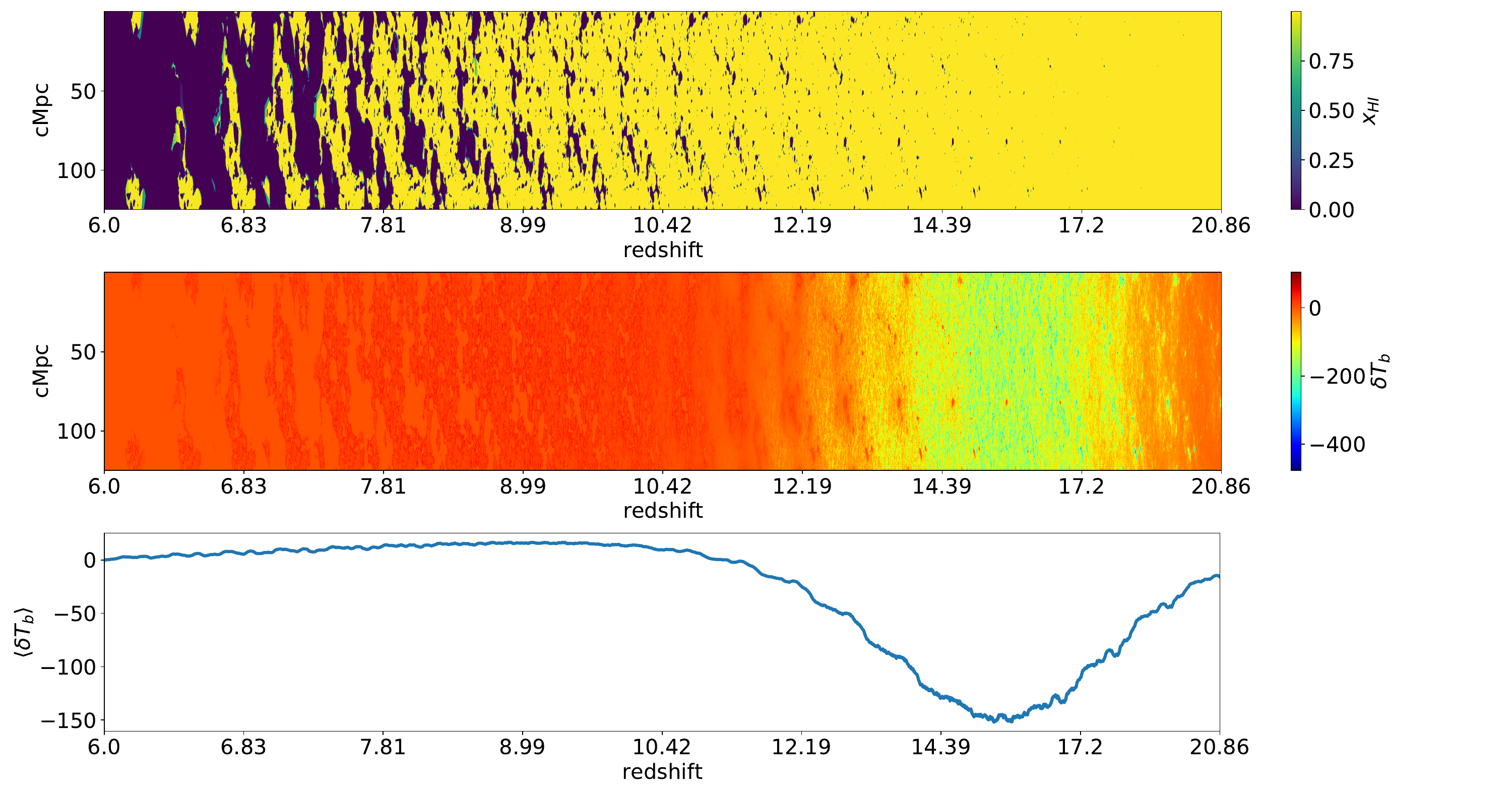}

\caption{\label{figure:lcslice} Demonstration of a slice in the redshift direction of one HIGH-RES simulation. The upper panel shows the evolution of the \(x_{\mathrm{HI}}\) (neutral hydrogen fraction). The middle panel displays the 21 cm differential brightness temperature. The lower panel presents the global 21 cm differential brightness temperature evolution.}
\end{figure*}

\begin{table*} 
 \centering

    \caption{Parameter descriptions of ionisation simulations.}
    \label{table: simulation information}
\begin{tabular}{p{4cm}|p{4cm}|p{5cm}|p{1cm}|p{3cm}}
\hline\hline
Name & ionisation simulation resolution & Number of light cones $\times$ (field of view) & $\zeta$ & $T_{\rm vir}(M_{\rm min})$ \\ 
   &(cMpc)& &  &  (K)  \\
\hline
Faint galaxies model& 1.5625    &$1\times (2.5^{\circ}\times 2.5^{\circ})$&20     &    $ 2\times10^4 $      \\
Bright galaxies model&1.5625   &$1\times (2.5^{\circ}\times 2.5^{\circ})$&200             &       $2\times10^5 $    \\
LOW-RES  &2    &$16\times (0.625^{\circ}\times 0.625^{\circ})$&30               &       $5\times10^4 $    \\
MID-RES  & 1    &$16\times (0.625^{\circ}\times 0.625^{\circ})$&30            &       $5\times10^4 $    \\
HIGH-RES  & 0.5    &$16\times (0.625^{\circ}\times 0.625^{\circ})$&30              &      $5\times10^4 $    \\

\hline
\hline
\end{tabular}
\end{table*}

\begin{figure*} 
\centering
\includegraphics[width=1\linewidth]{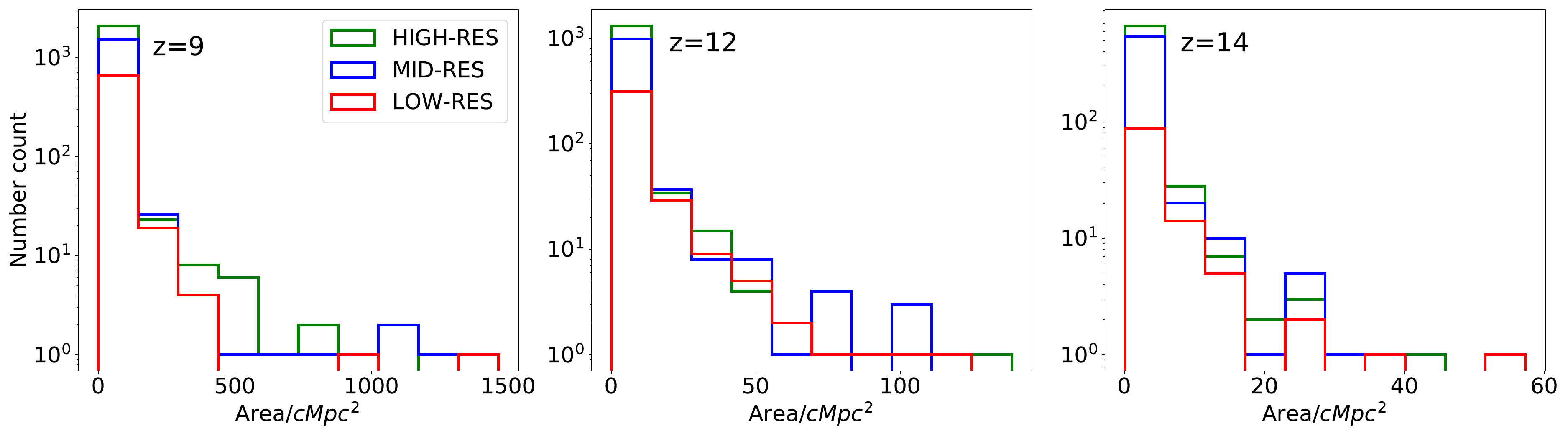}

\caption{\label{figure: source_number}  Number counts of ionised bubbles as a function of projected area of bubble for simulation slices at redshifts $z$\(= 9\) (left), $z$\(= 12\) (middle), and $z$\(= 14\) (right), respectively. Bubbles were identified via a Friends‑of‑Friends (FoF) algorithm \citep{2006MNRAS.369.1625I} using an ionisation fraction threshold of 0.5. The HIGH-RES resolves the highest abundance of small‑scale ionised regions across all three epochs, highlighting the impact of spatial resolution on the recovered small-scale ionisation topology.} 
\end{figure*}

\begin{figure*}
\centering
\includegraphics[width=1\linewidth]{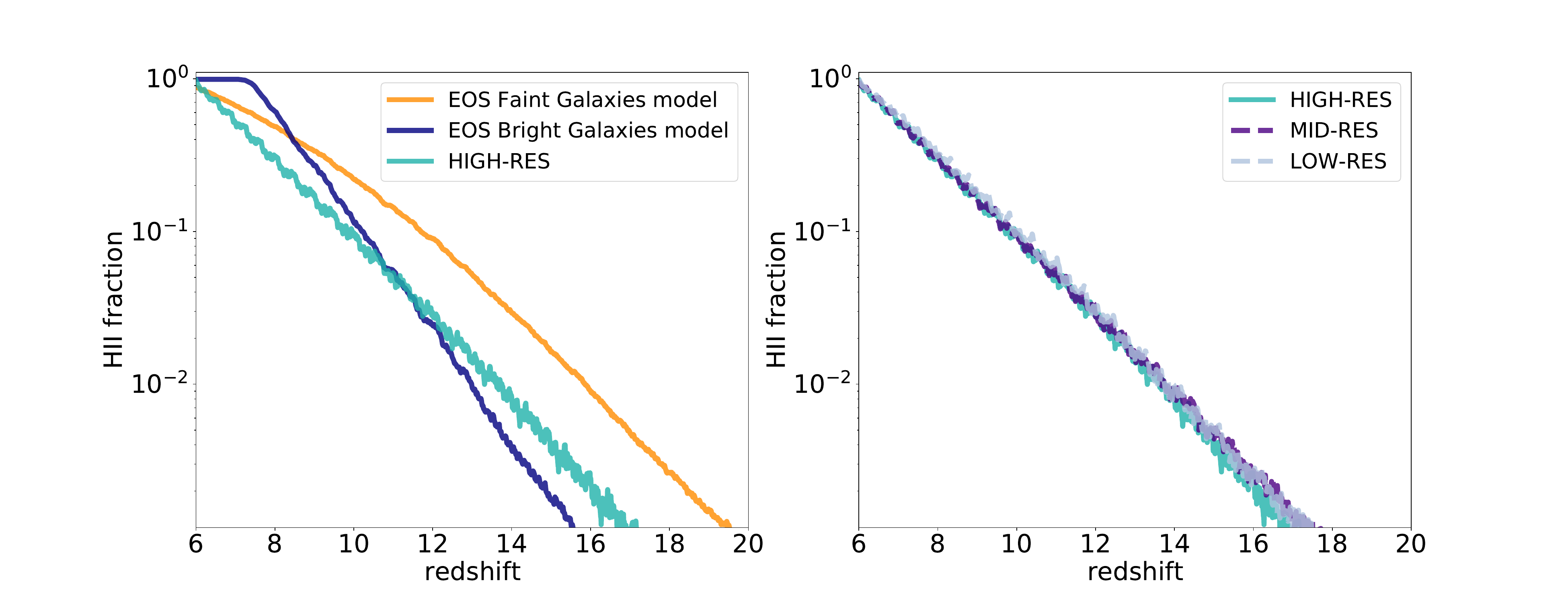}

\caption{\label{figure: HII versus redshift} Global volume-averaged HII (ionised hydrogen) fraction evolutions for the five ionisation models (listed in Table~\ref{table: simulation information}) used in this work. 
    \textbf{Left:} Ionised fraction evolution for three ionisation models with different ionisation parameter settings: EOS Faint Galaxies model, EOS Bright Galaxies model, and HIGH-RES. 
    \textbf{Right:} Ionised fraction evolution for three models with the same ionisation parameter settings but different resolution settings: HIGH-RES, MID-RES, and LOW-RES. 
    The different colours represent different models. 
    Left panel shows that simulations with different parameter settings exhibit distinct ionisation processes, such as variations in the reionisation redshift range and the speed of ionisation. 
    Conversely, the right panel shows that simulations with the same ionisation parameter settings (\(\zeta\) and \(M_{\rm min}\)) have similar reionisation processes, despite differences in ionisation simulation resolution.}
\end{figure*}

This study utilises five distinct reionisation simulations to systematically investigate the effects of ionisation parameter settings and simulation resolution. The selection of multiple simulations is motivated by two key considerations: one is the current lack of well-constrained ionisation parameters, and another is the need to explore resolution-dependent effects in ionisation simulations. Among the five simulations, two are adopted from the EOS project data release \citep{2016MNRAS.459.2342M}, specifically the Faint Galaxies and Bright Galaxies models. To complement these established models, we have developed three additional simulations with varying resolutions, designated as LOW-RES (low resolution), MID-RES (medium resolution), and HIGH-RES (high resolution), which were produced by \texttt{21CMFAST v3} \citep{2016MNRAS.459.2342M, Murray2020}. The complete specifications of all simulations are provided in Table~\ref{table: simulation information}. We note that the EoR simulations used in this work assume that foregrounds are perfectly removed.

We employ two contrasting reionisation models from the EOS project: the Faint Galaxies model and the Bright Galaxies model. These complementary approaches enable systematic investigation of ionisation model dependencies in gravitational lensing effects, handling the unresolved debate concerning the relative contributions of faint and bright galaxies to reionisation processes \citep{2019ApJ...879...36F, 2020ApJ...892..109N}. In the Faint Galaxies model, ionisation sources are dominated by small haloes near the atomic cooling threshold of \( T_{\rm vir} = 2 \times 10^4 \, \text{K} \), with an ionising efficiency of \(\zeta = 20\). Conversely, the Bright Galaxies model implements an elevated virial temperature threshold \( T_{\rm vir} = 2 \times 10^5 \, \text{K} \) to account for suppressed star formation in low-mass haloes through efficient supernova feedback mechanisms, while adopting a significantly enhanced ionising efficiency of \(\zeta = 200\). These simulations are processed into two light cones, each with a field of view of \(2.5^\circ \times 2.5^\circ\). Cosmological parameters used in these two EOS simulations are from the Planck 2015 \citep{2016A&A...594A..13P}, which are slightly different from the Planck 2018 Cosmological parameters adopted in other parts of the simulations in this work.

To systematically evaluate resolution-dependent effects in our analysis, we performed three additional ionisation simulations using the public semi-numerical code \texttt{21CMFAST v3} \citep{Murray2020}: LOW-RES, MID-RES, and HIGH-RES. These simulations share identical ionisation parameters (\(\zeta=30\), \(T_{\rm vir} = 5 \times 10^4 \, \text{K}\)) but employ progressively refined spatial resolutions: the ($125$ cMpc)$^3$ volume with initial density grids of $192^3, 384^3$, and $768^3$ voxels, subsequently smoothed down to ionisation maps with  $64^3, 128^3$, and $256^3$ grids having cell sizes of approximately 2, 1, and 0.5 cMpc respectively. For each resolution configuration, we generated sixteen \(0.625^\circ \times 0.625^\circ\) independent light cones using distinct initial density fields to mitigate sample noise from cosmic variance. As an illustration, Figure~\ref{figure:lcslice} shows a slice of one HIGH-RES light cone. For the simulation prescription, all three models are generated using \texttt{21CMFAST v3} with inhomogeneous recombinations \citep{2014MNRAS.440.1662S} enabled and without assuming a saturated spin temperature.

Our analysis reveals a resolution dependence in the characterisation of small-scale ionised bubbles across the LOW-RES, MID-RES, and HIGH-RES simulation suites. Comparison of ionised bubble counts in 2D slices at redshifts \(z=9 \), \(z=12 \), and \(z=14 \)  is shown in Figure~\ref{figure: source_number}, which demonstrates that higher spatial resolution enables the presence of significantly more small-scale ionised regions. Here we employ the Friends-of-Friends (FoF) algorithm \citep{2006MNRAS.369.1625I} using the \texttt{Tools21cm} package \citep{2020JOSS....5.2363G} for bubble identification, which provides one important methodological advantage for comparison that it directly shows the number of ionised bubbles in each size bin. The linking length of the FoF algorithm is set to the size of the corresponding original reionisation simulation grid.

The global ionisation evolution shows distinct patterns across different models (see Figure~\ref{figure: HII versus redshift}). While LOW-RES, MID-RES, and HIGH-RES simulations exhibit nearly identical reionisation histories due to shared parameter configurations (\(\zeta\) and \(T_{\rm vir}(M_{\rm min})\)), they differ substantially from both Faint and Bright Galaxies models.  Specifically, the elevated \(T_{\rm vir}(M_{\rm min})\) threshold in the LOW/MID/HIGH-RES ensemble postpones the onset of ionisation relative to the Faint Galaxies scenario, where smaller dark matter haloes facilitate earlier reionisation contributions. The Bright Galaxies model shows the most delayed ionisation, consequent to its high \(T_{\rm vir}(M_{\rm min})\). Furthermore, the augmented \(\zeta\) induces a contraction of the reionisation timescale (\(\Delta z_{\rm re}\)), resulting in intermediate \(\Delta z_{\rm re}\) values for the LOW/MID/HIGH-RES simulations that bridge the extended Faint Galaxies and rapid Bright Galaxies regimes.

\subsection{Models of foreground deflectors} \label{section: Foreground deflector simulation}

\begin{figure*}
\centering
\includegraphics[width=1\linewidth]{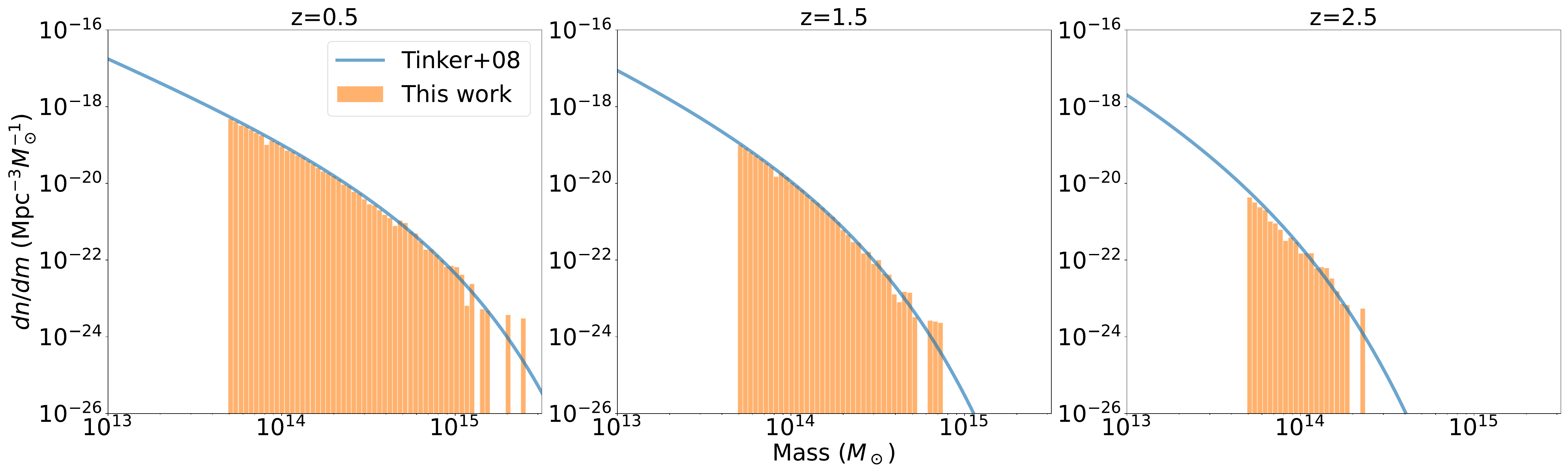}
\caption{Comparison between the mass functions of the deflectors in our light cones (red points) and the theoretical outcomes given by  \protect\cite{2008ApJ...688..709T} (blue lines) at redshifts $z$= \( 0.5\), \(1.5\), and \(2.5\), from left to right, respectively.}
\label{figure: crosscheck_mass_z}
\end{figure*}

\begin{figure*}
\centering
\includegraphics[width=1\linewidth]{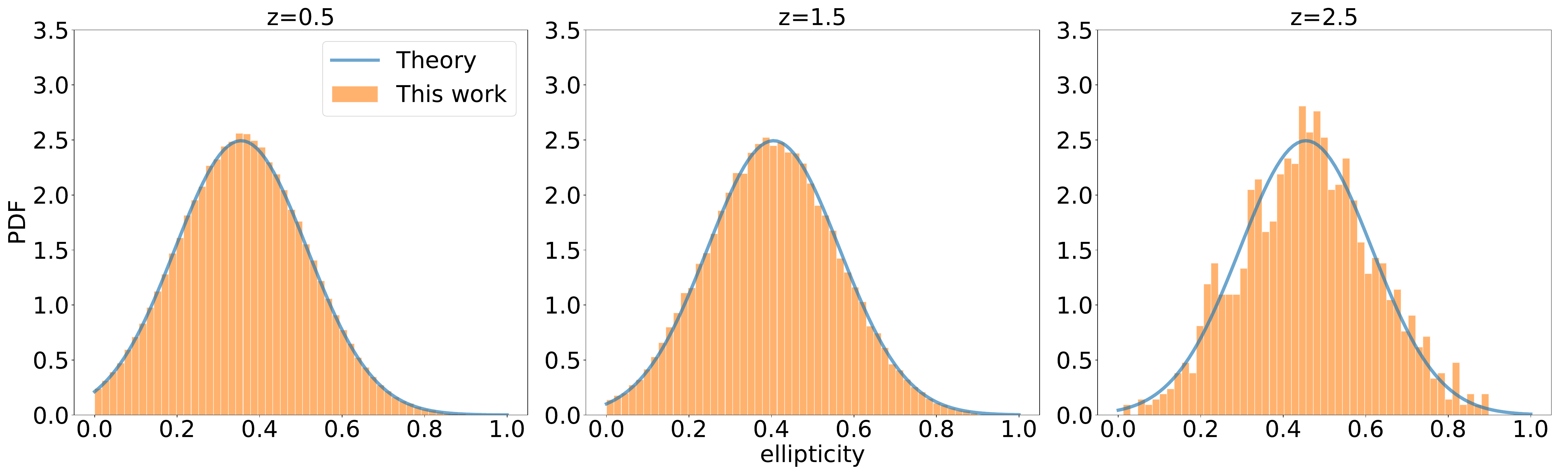}
\caption{Comparison between the PDFs of projected mass ellipticity $e$ of the deflectors in our light cone (orange bars) and theoretical outcomes of the mean $e$ evolution given by \protect\cite{2005ApJ...618....1H} with Gaussian dispersion \(\sigma_e=0.16\) (blue lines) at redshifts $z$= \(0.5\), \(1.5\), and \(2.5\), from left to right, respectively.}
\label{figure: crosscheck_e} 
\end{figure*}

We use the TNFW model \citep{1997ApJ...490..493N,2009JCAP...01..015B} to describe the mass profiles of the foreground deflectors, which is adopted broadly to model dark matter haloes \citep{2016ApJ...817...24M,2017ApJ...845..118M, Dai_2018} that dominate galaxy clusters. The TNFW profile analytically renders the cluster lenses' macro mass distribution and realistically reflects their primary features, like concentration and virial mass. The 3D TNFW model can be given by \citep{2009JCAP...01..015B}:
 \begin{equation}
     \rho(x)=\frac{\rho_{\mathrm{s}}}{x(1+x)^2}\frac{{r_{\mathrm{t}}}^2}{(xr_{\mathrm{s}})^2+{r_{\mathrm{t}}}^2}\label{eq: rho_tnfw},
 \end{equation}
 with $x=\theta {D_{\mathrm{d}}}/r_{\mathrm{s}}$, where $\theta$ is the angular position of $x$. $r_{\mathrm{s}}$ and $\rho_{\mathrm{s}}$ are the scale radius and the characteristic density of the halo. $r_{\mathrm{s}}$ is calculated by $r_{200}=r_{\mathrm{s}}c$ and $c$ is the concentration parameter, which is obtained via the mass-concentration relation \citep{2018ApJ...859...55C}. $r_{\mathrm{t}}$ is the truncation radius, on account of the best-fit value of $r_{\mathrm{t}}$ in \cite{10.1111/j.1365-2966.2011.18481.x}, we set $r_{\mathrm{t}} = 2.6 r_{200}$. The projected surface mass density of the TNFW model can be given by \citep{2009JCAP...01..015B}:
\begin{align}
\label{eq: 2ddensity}
\Sigma(x)=&\frac{2 r_s \rho_s \tau^2}{(\tau^2+1)^2}
\Bigg\{\frac{\tau^2+1}{x^2-1}\left[1-F(x)\right]+2 F(x)\nonumber\\
&-\frac{\pi}{\sqrt{\tau^2+x^2}}
+\frac{\tau^2-1}{\tau\sqrt{\tau^2+x^2}}\,L(x)\Bigg\},
\end{align}
where
\begin{equation}
L(x)=\ln\left(\frac{x}{\sqrt{\tau^2+x^2}+\tau}\right),
\end{equation}
and
\begin{equation}
F(x)=\frac{\cos^{-1}(1/x)}{\sqrt{x^2-1}},
\end{equation}
with $\tau=r_{\mathrm{t}}/r_s$. Besides, the potential $\psi$ can be expressed as \citep{2009JCAP...01..015B}:
\begin{align}
\label{eq: tnfwphi}
\psi(x)=&\frac{8\pi G\rho_{\mathrm{s}} r_{\mathrm{s}}^3}{c^2}\frac{1}{(\tau^2+1)^2}\Bigg\{2\tau^2\pi
\left[\tau-\sqrt{\tau^2+x^2}+\tau\,
\ln\left(\tau+\sqrt{\tau^2+x^2}\right)\right]\nonumber\\
&+2(\tau^2-1)\tau\sqrt{\tau^2+x^2}\,L(x)+
\tau^2(\tau^2-1)\,L^2(x)\nonumber\\
&+4\tau^2(x^2-1)F(x)+\tau^2(\tau^2-1)\left(\cos^{-1}
\frac{1}{x}\right)^2\nonumber\\
&+\tau^2\left[(\tau^2-1)\ln \tau-\tau^2-1\right]\ln{x^2}\nonumber\\
&-\tau^2\left[(\tau^2-1)\ln \tau\,\ln(4\tau)+2\ln(\tau/2)-2\tau(\tau-\pi)\ln(2\tau)\right]\Bigg\}.
\end{align}
The deflection angle \(\boldsymbol{\alpha}\) of the TNFW model can be derived from the gravitational potential \(\psi\) using Eq.~\eqref{eq:deflection}. The elliptical case of the TNFW can be derived from the above spherical counterparts \citep{2002A&A...390..821G, 2009JCAP...01..015B}. Therefore, each deflector in a given light cone can be characterized by six parameters: (R. A., Dec.) (angular position), $\phi$ (position angle of the projected major axis), $M_{200}$ (mass), $z$ (redshift), and \(e\) (ellipticity, defined as \(e = 1 - q\), where $q$ is the axis ratio). 

To build a light cone containing deflectors described by the TNFW profile, one can randomly resample the above combination of six parameters according to the assumed distributions. Here, R. A. and Dec. are randomly generated with respect to uniform distributions. Regarding other parameters, the Python package \texttt{HALOMOD} \citep{2021A&C....3600487M} is used to provide the theoretical halo mass function at different redshifts, then $M_{200}$ and $z$ of deflectors are generated. For the halo mass function, we use the form given by \cite{2008ApJ...688..709T}:
\begin{equation}
\frac{{\mathrm{d}}n}{{\mathrm{d}}M} = f(\sigma) \frac{\overline{\rho}_{\mathrm{m}}}{M} \frac{{\mathrm{d}}\ln \sigma^{-1}}{{\mathrm{d}}M}, \label{eq: dndm}
\end{equation}
where \(\overline{\rho}_{\mathrm{m}} = \Omega_{\rm m} \rho_{\mathrm{crit}}\), \(\Omega_{\rm m}\) is the total matter density parameter, and the halo mass is defined as \(M_{200}\). Specifically, \(M_{200} = \frac{800}{3} \rho_{\mathrm{crit}} \pi r_{200}^3\), with \(\rho_{\mathrm{crit}}\) representing the critical density. The function \(f(\sigma)\) is parameterized as:
\begin{equation}
f(\sigma) = A \left[ \left( \frac{\sigma}{b} \right)^{-a} + 1 \right] e^{-c / \sigma^2}, \label{eq: fsigma}
\end{equation}
where the parameters \(a\), \(b\), \(c\), and \(A\) are taken from \cite{2008ApJ...688..709T}, \(\sigma\) is the standard deviation of the linear density field smoothed on scale \(R\):
\begin{equation}
\sigma^2 = \int P(k) \hat{W}(kR) k^2 \, {\mathrm{d}}k. \label{eq: mass variance}
\end{equation}
Here, \(\hat{W}(kR)\) is the Fourier transform of the real-space top-hat filter function of radius \(R\). \(P(k)\) is the linear matter power spectrum as a function of wave number \(k\), in  \texttt{HALOMOD}, the fitting formula for the linear transfer function including baryons from \cite{1998ApJ...496..605E} is adopted. The low-mass threshold for the foreground deflectors is set to \(M_{200} = 5 \times 10^{13} M_{\odot}\), and the maximum redshift cut for deflectors is set to \(z_{\mathrm{max}} = 4\). Figure~\ref{figure: crosscheck_mass_z} presents the comparisons of the halo mass number density distributions extracted from our foreground light cones and the theoretical halo mass functions at redshifts \(z = 0.5\), \(1.5\), and \(2.5\), validating the resampling procedure. Besides, we draw the 2D projected ellipticity \(e\) of the mass distribution of haloes based on the formula in \citet{2005ApJ...618....1H}: the redshift-dependent mean value of \(e\) evolves as \(\mu_e(z)=0.33+0.05\,z\), which is obtained from their large-volume, high-resolution \(\Lambda\)CDM \(N\)-body simulations. At \(z\simeq0\), the mean value of \(e\) is in high agreement with the observation of the mass distribution of local Abell clusters \citep{1991MNRAS.249..662P}, local \(e\) distribution is well described in \cite{1991MNRAS.249..662P} by a Gaussian distribution with dispersion \(\sigma_e=0.16\). \cite{2006MNRAS.367.1781A} found that the 3D minor-to-major axis ratio of haloes up to $M_{vir}=2\times10^{14}M_{\odot}$ is well described by the Gaussian model with standard deviation $\sim 0.1$ over the redshift range $z$= 0 $\sim$ 3. \cite{Suto_2016} showed no significant difference ($\lesssim 15 \%$) of standard deviation of 2D minor-to-major axis ratio of haloes with $M_{vir}\gtrsim 6.25\times10^{13}M_{\odot}$ between $z$= 0 $\sim$ 1. These results are also obtained based on  \(N\)-body simulations. So we adopt $\sigma_e=0.16$ and fix it at all redshifts. The setting of $\sigma_e$ can be improved based on more data in the future. Figure~\ref{figure: crosscheck_e} shows the comparisons between the PDFs of the deflectors' ellipticities and the theoretical predictions at redshifts \(z = 0.5\), \(1.5\), and \(2.5\), validating our resampling procedure. At last, the projected position angles of deflectors $\phi$ are sampled following a uniform distribution.

\subsection{Lensing Simulation}\label{section: lensing simulation}

With the light cone of deflectors constructed using the approach described in Sect. \ref{section: Foreground deflector simulation} and the source light cone built upon the ionisation simulations specified in Table \ref{table: simulation information}, we performed multi-plane gravitational lensing simulations. This framework implements a ray-tracing algorithm that propagates light through the intervening matter distribution, generating lensed two-dimensional \(x_{\mathrm{HI}}\) maps at different redshifts.

First, to mitigate the Poisson noise inherent in Monte Carlo sampling, we generated an ensemble of 50 independent light cones of deflectors, named as \texttt{SET-FLC}, and each light cone covers a $2.5^{\circ} \times 2.5^{\circ}$ field of view. For comparative investigation, we adopt five source light cones: two of them are created on the EoR simulations with the assumptions of the EOS Faint Galaxies model and Bright Galaxies model separately, three of them are created on the HIGH/MID/LOW-RES simulations separately, see Sect. \ref{section: Reionization simulation generation}. For each of them, we perform 50 lensing simulations by connecting it to 50 light cones of deflectors from \texttt{SET-FLC}, and the final measurements of BSDs are the mean with $1 \sigma$ uncertainty of the outcomes of the MFP applied to all 50 realisations. Notably, the field of view of light cones of the HIGH/MID/LOW-RES simulations is one-sixteenth the size of the light cones in \texttt{SET-FLC}, measuring $0.625^{\circ}\times 0.625^{\circ}$. Hence, we equally divided each foreground light cone in \texttt{SET-FLC} into 16 subfields, each of which corresponds to one HIGH/MID/LOW-RES light cone for lensing simulations.

Second, we implement lensing simulations in the manner of the multi-lens-plane ray-tracing method using the \texttt{Lenstronomy} Python package \citep{2018PDU....22..189B, 2021JOSS....6.3283B}, and we place the haloes within the light cone onto a series of fixed lens planes at redshifts that are closest to them in terms of comoving distances. These lens planes span from $z=0.1$ up to $z=4$, with a redshift spacing of $\Delta z=0.1$ for each lens plane. Typically, there are $\sim 500$ haloes in a $2.5^{\circ} \times 2.5^{\circ}$ light cone. The angular resolution of our lensing simulation is 5 arcsec, and the discussion for verifying this choice is presented in Sect. \ref{section: uncertainties caused by lens light cone and ray tracing}. Deflection angle maps in all lens planes are calculated using the Python package \texttt{Lenstronomy} \citep{2018PDU....22..189B,2021JOSS....6.3283B}. On the $i$th lens plane, we also subtract the deflection angle caused by a constant mass sheet (denoted as $\kappa_{\rm{mean,i}}$) equivalent to the average surface density of the plane, so that ray tracing is driven by density fluctuations. $\kappa_{\rm{mean,i}}$ is the average convergence, uniformly sampled within the FOV range of the ray tracing map on the $i$th lens plane, and is expressed as:
\begin{equation}
\kappa_{\mathrm{mean},i}\equiv \langle \kappa_i(\theta_i)\rangle_{\mathrm{FOV}}\approx \frac{1}{N_{\mathrm{pix}}}\sum_{p=1}^{N_{\mathrm{pix}}}\kappa_i(\theta_p)
\end{equation}
where $N_{\mathrm{pix}}$ is the number of pixels uniformly sampled, $\kappa_i(\theta_i)$ is the convergence at angular position $\theta_i$ on the $i$th lens plane, which is calculated by Eq. \eqref{eq:kappa_on_eachplane}, and $\theta_p$ is the angular position of a sampling point on the uniform grid. The sampling resolution is set to be the same as the resolution of the ray tracing map, which is 5 arcsec in the fiducial run. The deflection angle caused by the mass sheet of $\kappa_{\rm{mean,i}}$ at angular position $\theta$ is given by \citep{2022iglp.book.....M}:
\begin{equation}
\mathbf{\alpha}(\mathbf{\theta})=\kappa_{\rm{mean,i}}\mathbf{\theta}.
\end{equation}
The unlensed ionisation maps are created using the same approach and configurations mentioned above, but no deflection angles are included. Therefore, the unlensed ionisation maps are essentially an upsampling of the primordial 2D \(x_{\mathrm{HI}}\), which maintains consistent resolution, theoretical and numerical assumptions when measuring the difference between lensed and unlensed BSDs.

\section{Estimation of BSD}\label{section: Measuring BSD}
\begin{figure}
    \centering
    \begin{minipage}[t]{1\linewidth}
        \centering
        \includegraphics[width=1\linewidth]{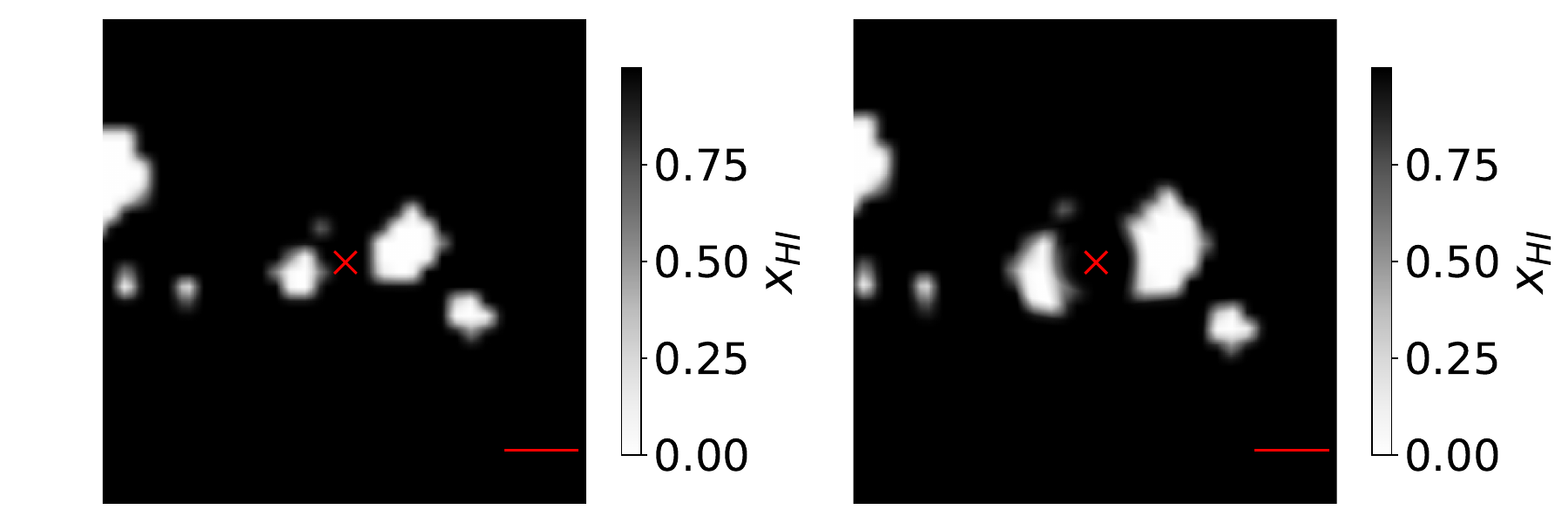}
      
    \end{minipage}

    \caption{\label{figure: showcases of distorted bubble} A showcase about how lensing changes the morphology of the ionised bubble of HIGH-RES at redshift $z = 12$. The panels from left to right are an unlensed $x_{\rm HI}$ map and the corresponding lensed one. The red segments at the bottom-right of each panel correspond to 50 arcsec, and the oblique cross symbols in the centre indicate the positions of foreground deflectors. This comparison shows that lensing significantly distorts the morphology of the ionised bubbles.}
 \end{figure}

\begin{figure}
\centering
\includegraphics[width=0.8\linewidth]{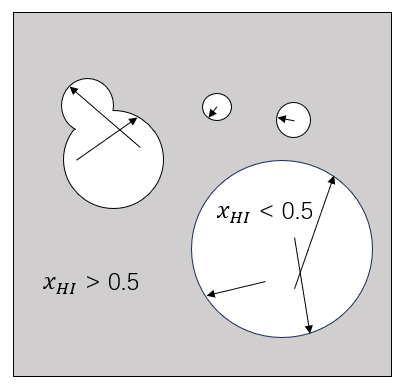}

\caption{\label{figure: mfp} A schematic diagram of using the MFP method to measure the BSD. This picture is a cartoon diagram of a two-dimensional $x_{{\mathrm{HI}}}$ map during EoR, we randomly select starting points in ionised regions of this map where $x_{{\mathrm{HI}}}$ < 0.5 (white regions) and shoot rays in random directions until the rays hit the edge of the un-ionised region (dark-coloured region, $x_{{\mathrm{HI}}}$ > 0.5) or the boundaries of map. The length of the ray's path is used to characterise the size of the ionised region. By repeating this process, we can obtain the probability distribution of ionised regions' size across the entire $x_{{\mathrm{HI}}}$ map.
}
\end{figure}

\begin{figure*}
    \centering
    \begin{minipage}[t]{1\linewidth}
        \centering
        \includegraphics[width=1\linewidth]{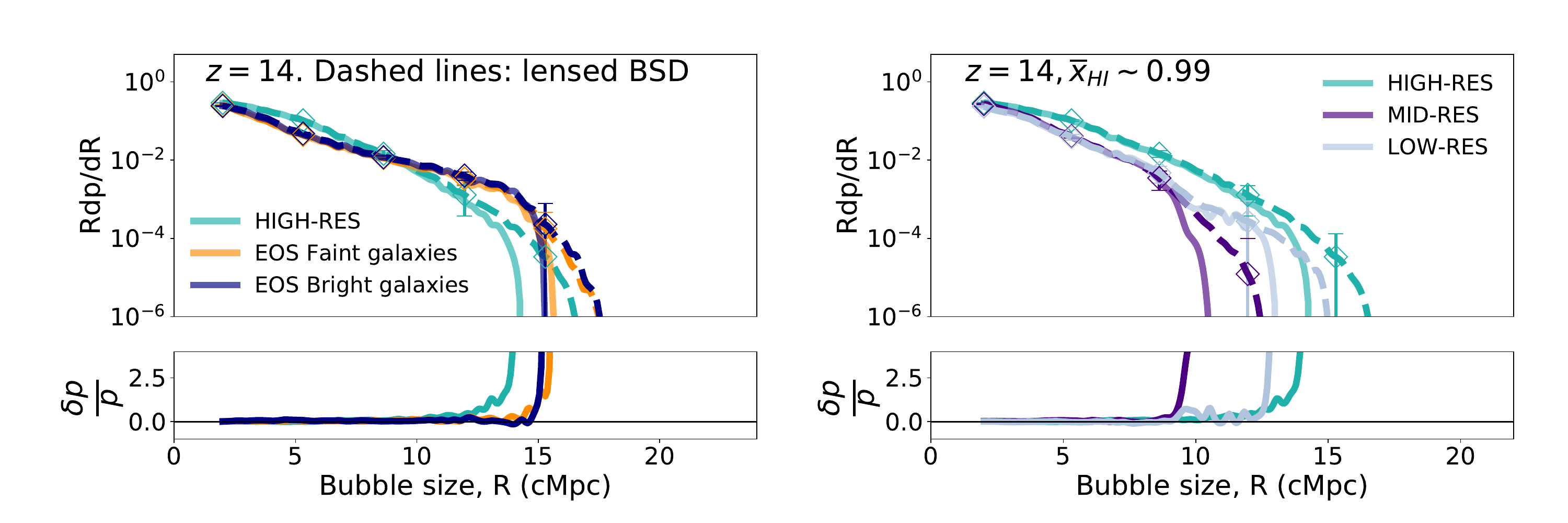}
      
    \end{minipage}
    \begin{minipage}[t]{1\linewidth}
        \centering
        \includegraphics[width=1\linewidth]{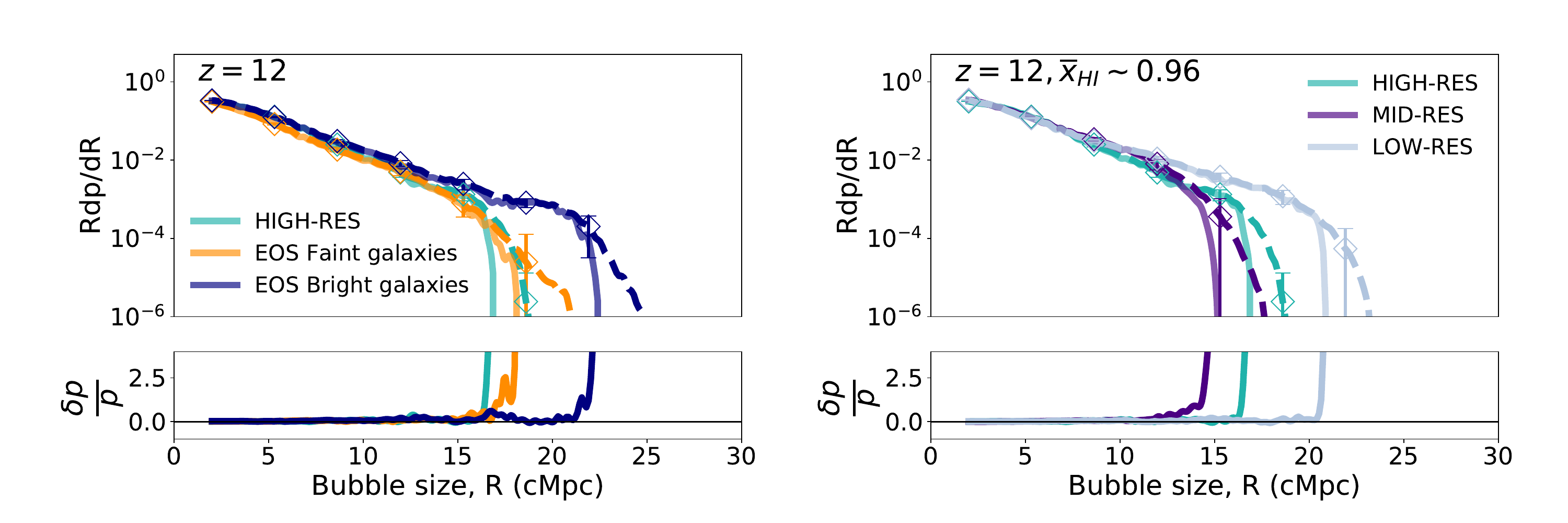}
    
    \end{minipage}

    \begin{minipage}[t]{1\linewidth}
        \includegraphics[width=0.5\linewidth]{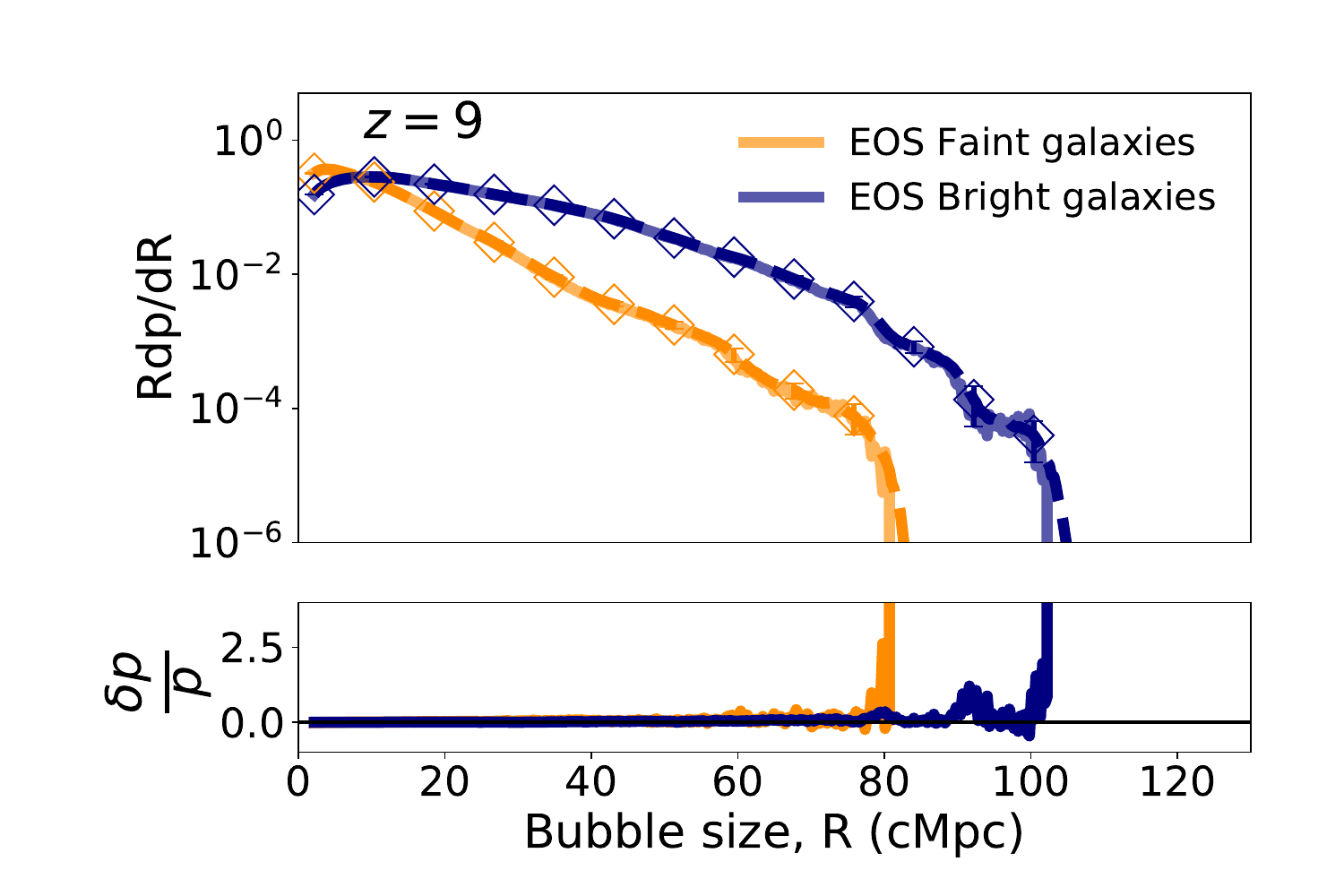}
    
    \end{minipage}
    \caption{\label{figure: bubble_size_pdf} Bubble Size Distribution (BSD) for lensed and unlensed cases at redshifts \(z = 14\), \(z = 12\), and \(z = 9\), from top to bottom. Solid lines indicate unlensed BSD and dashed lines represent lensed BSD. Different colours represent the results of different ionisation models. The left panels of $z$ = 12 and $z$ = 14 compare the results of three different ionisation models which have different ionisation parameter settings (\(\zeta\) and \(T_{\rm vir}(M_{\rm min})\)), while the right panels compare results of three simulations with the same ionisation parameters but different ionisation simulation resolutions. The error bars on the lensed curves correspond to 1$\sigma$ uncertainties in the 50 lens light cone results. At the small R end, the error bars are hardly visible because the uncertainties are small. Below each chart, we show the relative difference between the lensed and unlensed BSD, given by $\delta p/p$=(PDF$^{\prime}$-PDF)/PDF where PDF$^{\prime}$ is the bubble size probability distribution function under the lensed situation, and PDF is that of the unlensed situation. Besides, the slice thickness used for BSD measuring for each model is fixed to 2 cMpc here.}
 \end{figure*}

The results of the lensing simulation show that gravitational lensing alters the morphology of ionised bubbles. For example, as shown in Figure \ref{figure: showcases of distorted bubble}, the shapes of the two ionised bubbles in the central region are distorted by foreground lensing. This distortion extends beyond mere visual changes; it also impacts the statistical properties of ionised bubbles, such as the BSD. To characterise this lensing-induced modification, we require a robust BSD quantification methodology. The present section details our approach for accurate BSD measurement.

We apply the MFP method \citep{2007ApJ...669..663M} to measure the BSD from the 2D unlensed $x_{{\mathrm{HI}}}$ map and the lensed $x_{{\mathrm{HI}}}$ map. We note that the BSD referred to in this paper is apparent BSD. By comparing the BSD before and after lensing, we were able to quantify the impact of lensing on the BSD. The MFP method is one of the most accurate approaches for BSD measurements compared to other methods in the literature, as demonstrated in \cite{2016MNRAS.461.3361L}, it performs exceptionally well in recovering the intrinsic BSD. This method is based on a Monte Carlo algorithm. We randomly select points within the ionised regions on the two-dimensional $x_{{\mathrm{HI}}}$ map and shoot rays in random directions. The rays stop when they encounter the boundary of the un-ionised region, which is defined as the area where the ionisation fraction is less than 0.5, or stop when they encounter the boundary of the map. The length of the path travelled by the ray is used to characterise the size of the ionised region (as shown in Figure \ref{figure: mfp}). Continuously repeating this process allows us to determine the probability distribution of ionised bubble size. In this work, we use the Python package \texttt{Tools21cm} \citep{2020JOSS....5.2363G} to perform the MFP method to generate BSD for each ionisation model. For $z_s=$12 and 14, the operation of randomly shooting rays is repeated $10^6$ times. At $z_s=$9, the ionized region is much larger than that of $z_s=$12 and 14, we set the number of randomly shooting rays to be $10^8$, and we discuss the influence of the iteration number in Sect. \ref{section: uncertainties caused by BSD measuring}. Furthermore, the thickness of the 2D slices used to measure the BSD in MFP method is uniformly set to $2\,\mathrm{cMpc}$. In this process, we generate the slices with a thickness of $2\,\mathrm{cMpc}$ using linear interpolation; such a slice incorporates information from multiple original ionization simulation slices. Besides, to obtain robust results, we averaged the BSD measurements over the 50 lensing simulation realisations.

\section{Results}\label{section: results}

This section presents the lensing effects on the BSD during EoR and the influence of simulation setups on the lensing effects on BSD, such as the construction of the source light cone (Sect. \ref{section: uncertainties caused source light cone}), the construction of lens light cone and ray‐tracing simulation (Sect. \ref{section: uncertainties caused by lens light cone and ray tracing}), and the MFP algorithm used to extract the BSDs (Sect. \ref{section: uncertainties caused by BSD measuring}). 

\subsection{Lensing Effects on BSD}\label{section: effect on bubble size}

Gravitational lensing induces significant distortions in the BSD in the large bubble regime, particularly during the early stage of the EoR ($x_{\rm{HII}} \lesssim 0.05$). These distortions manifest as an enhancement in the expected abundance of large bubbles across all source ionisation models, persisting over an extended timescale corresponding to a redshift interval of $\Delta z \gtrsim 2$ (see Figure \ref{figure: bubble_size_pdf}). Specifically, for the EOS Faint Galaxies model, we observe a 219\% increase in the abundance of bubbles with bubble size R $\textgreater 15$ cMpc at redshift $z_s$ = 14, while the EOS Bright Galaxies model shows a more pronounced enhancement of 832\%. 

Besides, the effects of lensing on BSD evolve along redshifts, e.g., results from HIGH-RES simulations demonstrate that at $z_s$ = 14 with $Rdp/dR = 10^{-6}$, the bubble sizes increase by approximately 15\%, while at $z_s = 12$, the increase remains significant at 8\%. However, as reionisation progresses, the larger size of the ionised bubbles compared to the lensing cross-section of galaxy clusters in the later stage of EoR makes their properties less sensitive to distortion, such as the lensing effects on BSDs are suppressed at $z_s$ = 9 (see Figure \ref{figure: bubble_size_pdf}). 

For small-scale bubbles, the BSDs remain unaffected by lensing throughout the entire EoR across all ionisation simulations down to 2 cMpc. This defines the nominal lower bound of the BSD measurement, and the lensing effect on bubbles below it remains mysterious. It will be discussed in Sect. \ref{section: uncertainties caused by BSD measuring} and will be investigated in detail in the future.

\subsection{Preliminary test of anisotropy in lensing-induced distortion}




Since lensing is expected to strongly affect information in the directions perpendicular to the LOS direction, while having little impact in the radial direction, i.e., the LOS direction, we further perform a simple test of the anisotropy of the lensing-induced ionised region size distortion. The test results show that, if we discuss only the properties of ionised bubbles along the LOS direction, the corresponding statistics are expected to be much less affected by lensing. This is important because it suggests a possible way to recover the unlensed bubble statistics and deserves further study.

Our test method is as follows. We use the 3D simulation data spanning 100 cMpc along the redshift direction, centred at redshift 12. For the unlensed case, we use the 16 original unlensed HIGH-RES light-cone realisations. For the lensed case, we randomly select five lensing realisations from the full set of 50 and apply them to the same 16 original HIGH-RES light-cone realisations. This allows us to measure the projected MFP statistics directly in the three-dimensional light-cone data. Specifically, following the MFP method used in our main pipeline, we launch rays from random locations within ionised regions in random directions and terminate each ray when it first encounters a non-ionised region. We then measure the projected length of each MFP ray segment along the $x$, $y$, and $z$ directions. For an MFP ray segment
\begin{equation}
\boldsymbol{\ell}=L_{\rm MFP}\hat{\boldsymbol{n}},
\end{equation}
the projected length along direction $i$ is defined as
\begin{equation}
L_{{\rm proj},i}
=
\left|\boldsymbol{\ell}\cdot\hat{\boldsymbol{e}}_i\right|
=
L_{\rm MFP}
\left|\hat{\boldsymbol{n}}\cdot\hat{\boldsymbol{e}}_i\right|,
\qquad i=x,y,z,
\end{equation}
where $L_{\rm MFP}$ is the full MFP ray-segment length, $\hat{\boldsymbol{n}}$ is the unit direction vector of the ray, and $\hat{\boldsymbol{e}}_i$ is the unit vector along direction $i$. The $z$ direction corresponds to the LOS direction, while the $x$ and $y$ directions are transverse to the LOS. We measure the results for both the unlensed and lensed cases.

\begin{figure*}
    \centering
    \begin{minipage}[t]{1\linewidth}
        \centering
        \includegraphics[width=1\linewidth]{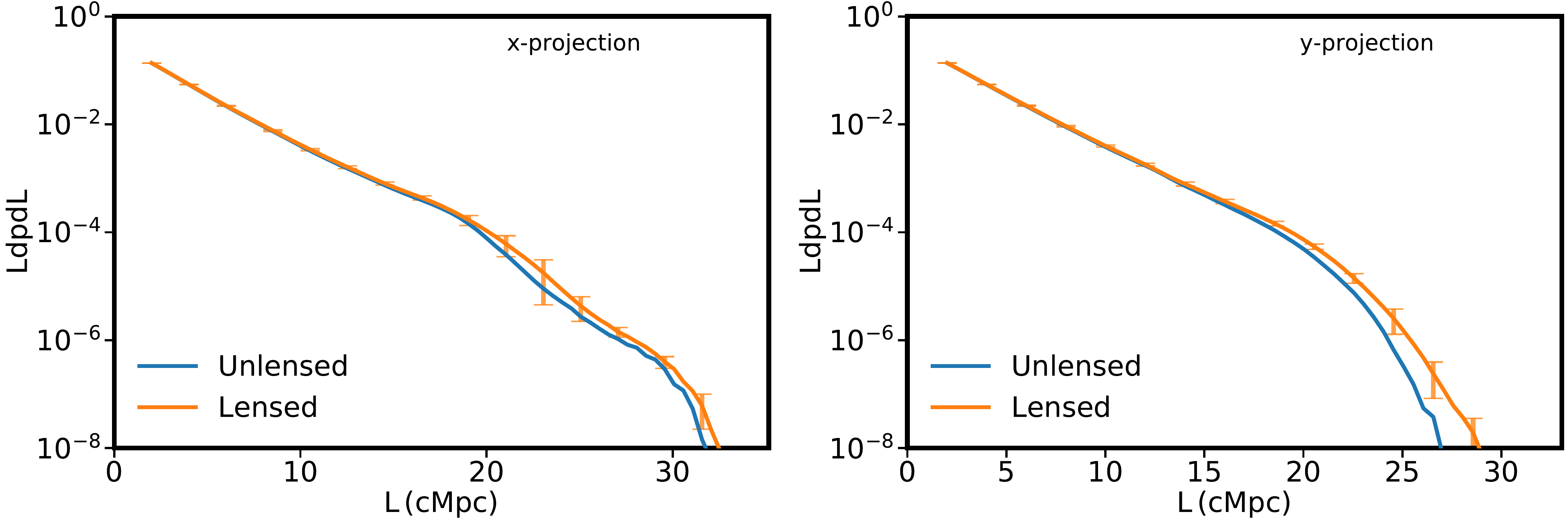}
      
    \end{minipage}
    \begin{minipage}[t]{1\linewidth}
        \centering
        \includegraphics[width=1\linewidth]{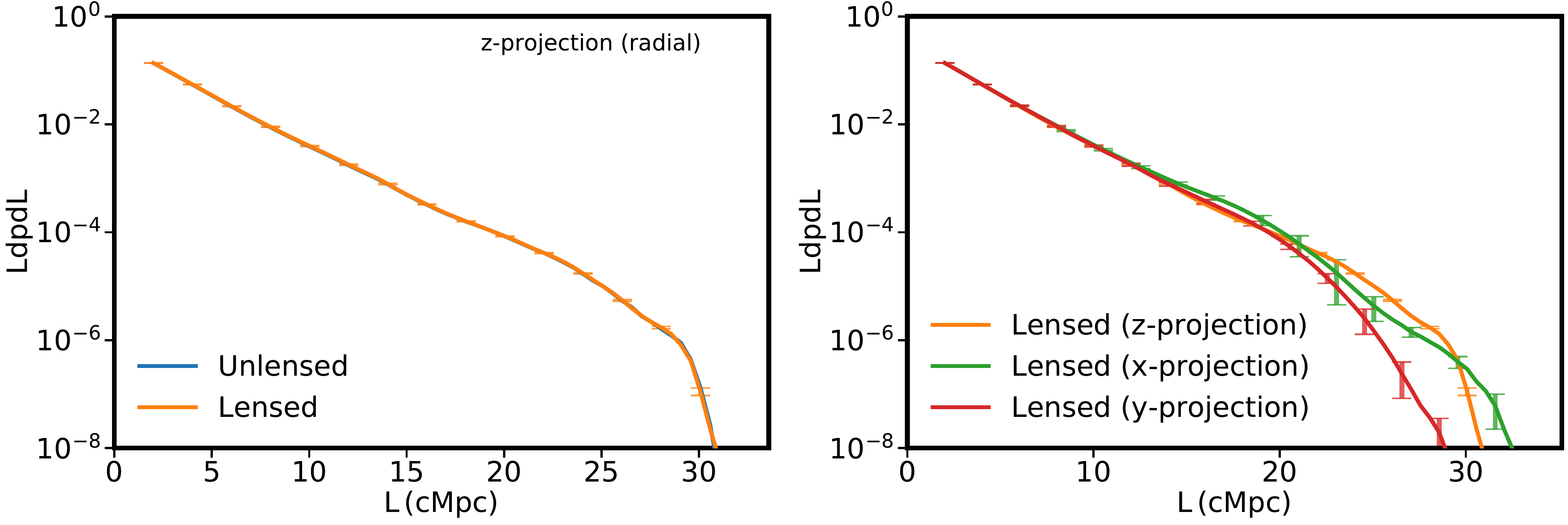}
    
    \end{minipage}

    \caption{Projected MFP ray-segment length distributions measured from the $z=12$ HIGH-RES simulation 3D data. The first row shows the unlensed and lensed results for the $x$- and $y$-projected lengths. In the lower row, the left-hand panel shows the unlensed and lensed results for the $z$-projected (radial or LOS-projected) lengths, while the right-hand panel compares the lensed results projected along the $x$, $y$, and $z$ directions. The error bars indicate the $1\sigma$ scatter among the five lensing realisations and are shown only at a subset of points for clarity.}
\label{figure: fixdi} 
 \end{figure*}

The test results are shown in Figure \ref{figure: fixdi}. We find that when the MFP ray-segment lengths are projected onto the LOS direction, the lensing effect is much weaker; when the ray-segment lengths are projected onto the $x$ or $y$ direction within the sky plane, the lensing effect becomes significantly stronger, similar to the results shown in Figure \ref{figure: bubble_size_pdf}.

In addition, the lower-right panel of Figure \ref{figure: fixdi} compares the three lensed projected-MFP distributions along the $x$, $y$, and $z$ directions. The three distributions are similar at the small-length end, while their differences become more apparent at the large-length end. We also tested the Poisson and jackknife uncertainties of these results. The Poisson uncertainties are small over the displayed range and have little impact on the results. By contrast, the jackknife uncertainties increase toward the large-length end, reflecting the greater sensitivity of the rare long projected MFP ray segments to the finite simulation volume. Therefore, the precise ordering of the three lensed distributions in this regime should not be over-interpreted, while the qualitative lensed–unlensed differences within each direction remain robust.

\subsection{Influence of Source Models on the Estimation of BSD} \label{section: uncertainties caused source light cone}

\begin{figure*}
\centering

\includegraphics[width=1.\linewidth]{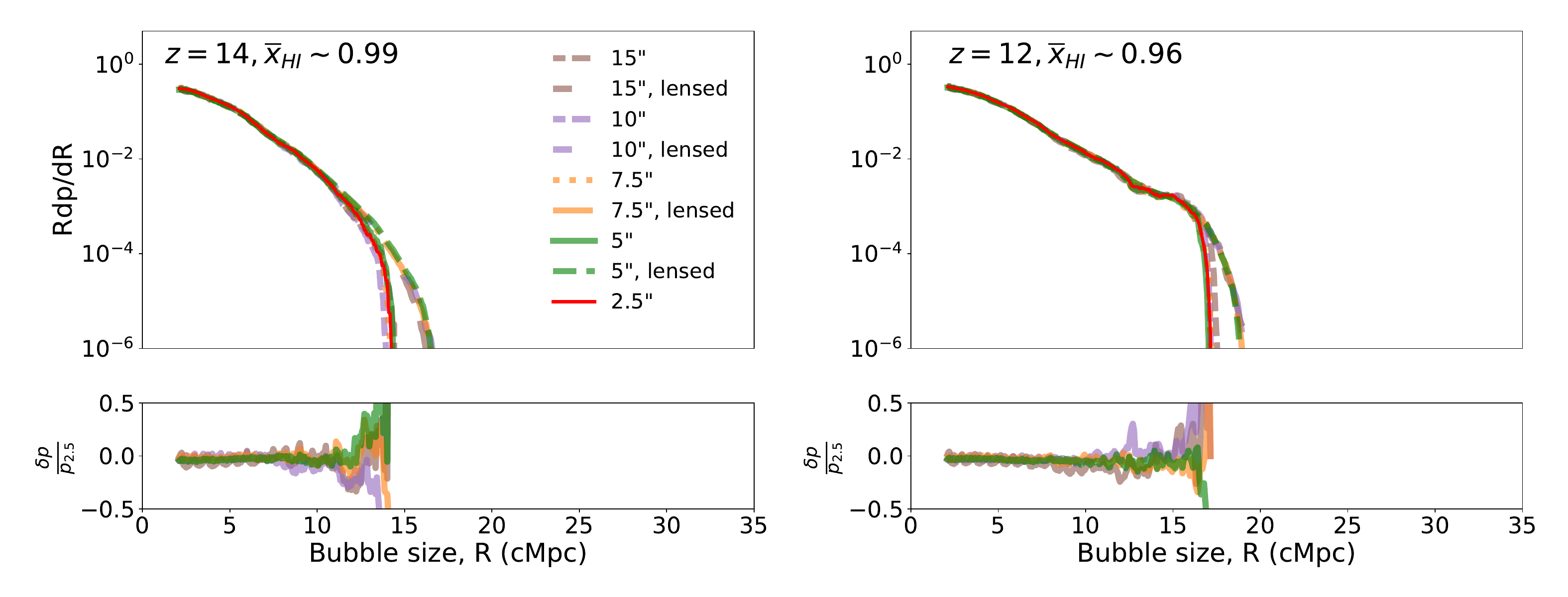}

\caption{
The influence of ray-tracing resolution on BSD. These panels show lensed and unlensed BSD of HIGH-RES when ray-tracing resolution is set to 5, 7.5, 10, and 15 arcsec at redshifts 12 and 14, respectively. We also show the unlensed BSD when the resolution is 2.5 arcsec as the red solid lines. BSD is measured by the MFP method \citep{2007ApJ...669..663M}. The lower panel shows the unlensed BSD difference compared to the case of 2.5 arcsec, given by $\delta p/p_{2.5}$=(PDF$^{*}$-$p_{2.5}$)/$p_{2.5}$. Where PDF$^{*}$ is the bubble size probability distribution function under different lensing simulation resolutions, and $p_{2.5}$ refers to the bubble size probability distribution of the case of 2.5 arcsec.}
\label{figure: test_BSD_resolution} 
\end{figure*}

BSDs are intrinsically sensitive to ionisation parameters, which might lead to various influences of lensing on BSDs. For example, the ionisation parameters minimum halo mass which host ionising sources \(M_{\rm min}\) and ionising efficiency \(\zeta\), where, increasing \(M_{\rm min}\) shifts the BSD toward larger bubbles, whereas a larger \(\zeta\) accelerates the growth of ionised bubbles \citep{2016MNRAS.459.2342M}. Expressly, at $z_s$ = 9, compared to the BSD of the EOS Faint Galaxies model, the BSD of EOS Bright Galaxies model exhibits more pronounced large ionised bubbles distribution for its greater \(M_{\rm min}\) as shown in the bottom-left panel in Figure \ref{figure: bubble_size_pdf}, which aligns with the difference between the simulation results of bright galaxies-dominated model and the faint galaxies-dominated model in \cite{2024MNRAS.528.4872L}. 

Besides, the Poisson uncertainty may affect the robustness of the measurements of lensing effects on BSDs, which is initially tied to the abundance of the corresponding dark matter haloes \citep{2011MNRAS.411..955M}. Poisson uncertainty on the lensing effects on BSDs evolves with the reionisation stage and the bubble scale. For example, when it comes to large bubbles (R $\gtrsim$ 10 cMpc) at $z$= 12, the difference between intrinsic BSDs of HIGH/MID/LOW-RES source models (Figure \ref{figure: bubble_size_pdf}) is largely due to the significant fractional Poisson uncertainty caused by the rarity of bubbles. To better quantify the lensing effects on BSDs, the impacts of Poisson uncertainty need to be improved by employing large-scale simulations with sufficient statistics in future studies.

Moreover, \cite{2007ApJ...669..663M} showed that a box size of 100 cMpc would underestimate the size of bubbles larger than the characteristic bubble size at $z_s$ = 9. Along with the progress of reionisation, more large bubbles are 'missed' due to the limited box size. Hence, the larger the ionised fraction, the larger the reionisation simulation box required to capture the ionisation topology (see Figure 7 and Figure 10 in \cite{2007ApJ...669..663M}). At redshifts above 12, the ionisation fraction is less than 0.05, and bubble sizes are approximately 0 to 25 cMpc, which is far smaller than the box size of source simulations ($\lesssim$ 0.06 $\times$ the size of EOS simulation and $\lesssim$ 0.2 $\times$ the size of HIGH/MID/LOW-RES simulations\footnote{For the HIGH-RES, MID-RES, and LOW-RES simulations, the light cone extends approximately 100 to 125 cMpc perpendicular to the LOS at different redshifts, while for the EOS simulations, it is approximately 400 to 500 cMpc.}). However, at redshift 9, with a larger ionisation fraction of about 0.2 and bubble sizes reaching 60 to 80 cMpc, the limited box size could impact BSD measurements in HIGH/MID/LOW-RES. In the latter half stages of reionisation ($x_{\rm{HI}}$ < 0.5), the size of large ionised bubbles exceeds 125 cMpc \citep{2024MNRAS.528.4872L}, causing the limited scales of HIGH/MID/LOW-RES simulations to have no chance in capturing the information of maximum bubble sizes. In this work, we only retain BSD results when the simulation box size is at least five times larger than the size of the largest bubble size, which relieves the effect of a finite box volume in ionised bubble size estimation. Large-scale simulations can address this limitation as well.

In addition, across all models and parameter choices tested, lensing by galaxy clusters affects the BSD in the large-size region because of its significant cutoff, which is a physical effect rather than a numerical assumption. For instance, the ionisation-field generation in \texttt{21cmFAST} is based on an excursion-set algorithm within the extended Press--Schechter formalism \citep{2004ApJ...613....1F}; this semi-numeric approach reproduces key properties of ionised regions, such as their topology and power spectra, in good agreement with radiative-transfer simulations \citep{2007ApJ...669..663M,2016MNRAS.461.3361L}, and has therefore been widely adopted. The excursion-set formalism predicts an exponential cutoff in the BSD at the large size end \citep{2004ApJ...613....1F}. The correlation between the cutoff positions and combinations of reionisation parameters suggests that information about the BSD cutoff provides a useful handle for constraining reionisation physics. Hence, using \textsc{21cmFAST}, we reran reionisation simulations with identical simulation seeds to test how this cutoff depends on $M_{\rm min}$ and $\zeta$, finding at $z=12$ and $z=14$ that decreasing $M_{\rm min}$ tends to shift the cutoff to larger $R$, plausibly because a lower threshold increases the abundance of ionising sources, triggering earlier reionisation and facilitating more frequent and efficient mergers of ionised regions. We also find that increasing $\zeta$ tends to move the cutoff to larger $R$, as expected from the enhanced ionising-photon budget. Tests based on three independent comparison experiments with different random seeds all support the above results, demonstrating a clear dependence of the cutoff scale on the reionisation parameters. Owing to the limited simulation volume and the vast parameter space, we restrict ourselves here to a set of simple tests and exploratory analyses; a statistically comprehensive quantification of how reionisation parameters and lensing jointly impact the BSD cutoff will be investigated in future work.

\subsection{Influence of the Setups of Lensing Simulations on the Estimation of BSD} \label{section: uncertainties caused by lens light cone and ray tracing}

Assumptions adopted in lensing simulation might introduce uncertainties and bias, which can not be ignored. First, we assume a uniform spatial distribution of the deflectors, neglecting the clustering of haloes and filamentary structures on cosmic scales. Second, our deflector mass profile is a single primary halo per deflector, while observational evidence suggests that accurate mass reconstructions of a galaxy cluster sometimes require multiple subhaloes, additional primary haloes, or external shear \citep{keeton2001computationalmethodsgravitationallensing,2007ApJ...662..781R,2016A&A...587A..80C,2020ApJS..247...12S}. As the presence of LOS haloes can increase the probability of lensing events \citep{2019ApJ...878..122L}, the effect of ignoring extra haloes along the LOS leads to an underestimation of the abundance of lensing events, thereby underestimating the impact of lensing on the BSD. Moreover, apart from the TNFW model, the mass profile of galaxy clusters can also be described by other lens models (such as the Pseudo Isothermal Elliptical Mass Distribution model \citep{1993ApJ...417..450K, elíasdóttir2007mattermergingclusterabell, Sharon_2020}), which gives various outcomes potentially. Above all, our assumptions lead to a lower limit of lensing effects on the BSDs. To better quantify the lensing effects on BSDs, we may model lensing light cones with cosmological simulations to involve (\romannumeral 1) more realistic spatial distributions of deflectors, and (\romannumeral 2) more comprehensive mass maps of deflectors.

Furthermore, numerical setups for the lensing simulation may also raise uncertainties and bias, such as insufficient ray-tracing resolution in lensing simulations, which smooths small-scale structures of lenses and sources, influencing the robustness of the results. To avoid this issue, we examine resolution dependence by testing values of 5, 7.5, 10, and 15 arcsec for the HIGH-RES model, measuring both lensed and unlensed BSDs at redshifts $z_s$ = 12 and 14 (Figure \ref{figure: test_BSD_resolution}). For convergence testing, we include an additional high-resolution case (resolution = 2.5 arcsec, shown as the red solid lines for unlensed BSD in Figure \ref{figure: test_BSD_resolution}). The results reveal diminishing differences in BSD values with decreasing resolution, with close agreement between the outcomes in the cases with angular resolutions of 2.5 and 5 arcsec. The relative difference of these two cases is $< 10\%$ except for the largest size end. Therefore, we use 5 arcsec as the angular resolution in our lensing simulation.

\subsection{Influence of the Setups of MFP on the Estimation of BSD} \label{section: uncertainties caused by BSD measuring}

\begin{figure*}
\centering
\includegraphics[width=1.\linewidth]{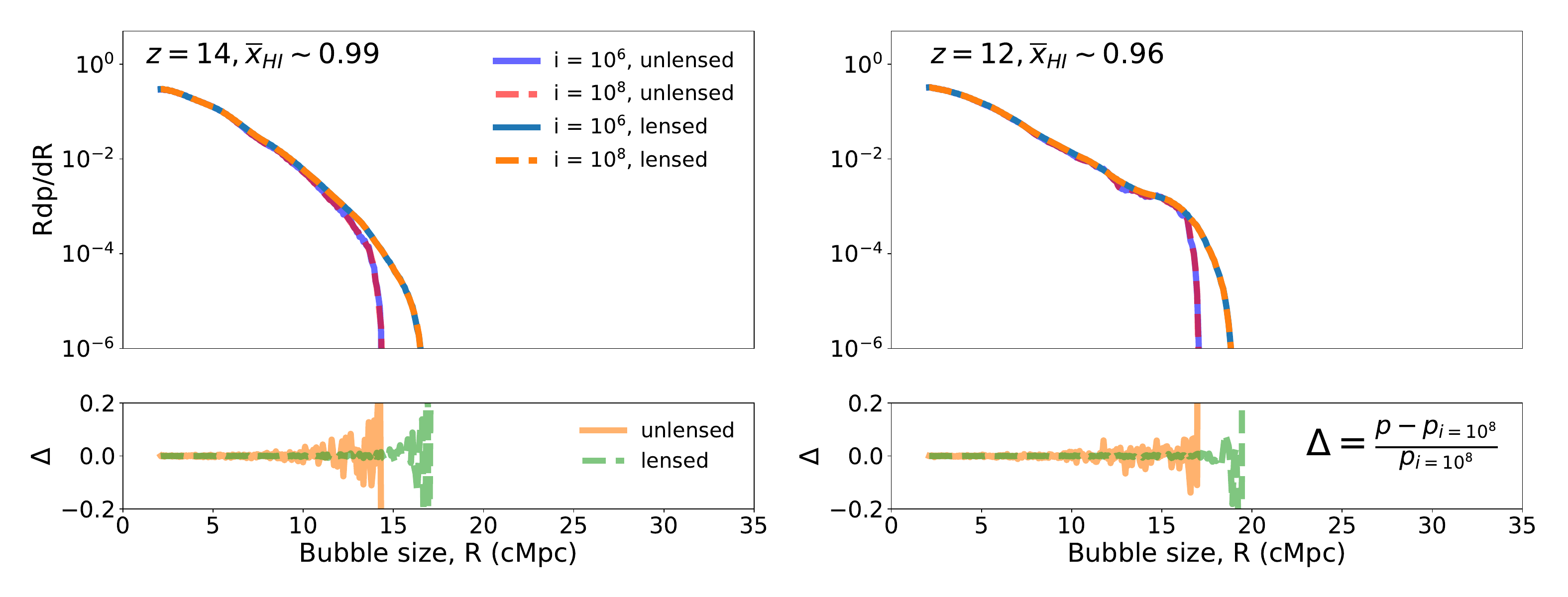}

\caption{
Robustness test of iteration counts in the MFP method for BSD measurement at $z_s = 14$ and $z_s = 12$ for HIGH-RES model. The upper panels present the standard implementation with $i=10^6$ iterations in the MFP algorithm (solid curves) and the realisation with $i=10^8$ iterations (dashed curves) for comparison. The lower panels show the relative BSD difference between realisations with $i=10^6$ and $i=10^8$, respectively. Left and right panels demonstrate the comparison when the source planes are located at $z_s = 14$ and $z_s = 12$, respectively.}
\label{figure: test_BSD_iterationnumber} 
\end{figure*}

\begin{figure*}
\centering
\includegraphics[width=1.\linewidth]{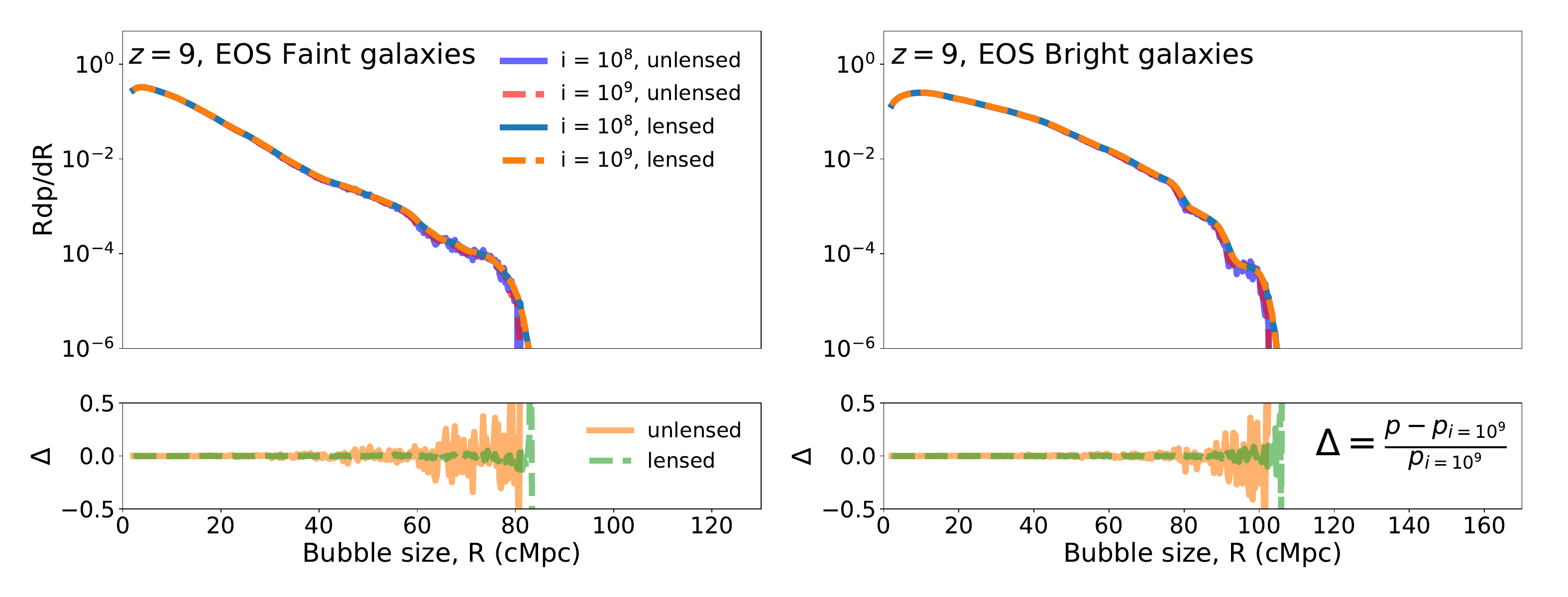}

\caption{
Robustness test of iteration counts in the MFP method for BSD measurement at $z_s = 9$ for the EOS Bright galaxies and Faint galaxies model. The upper panels present the standard implementation with $i=10^8$ iterations in the MFP algorithm (solid curves) and the realisation with $i=10^9$ iterations (dashed curves) for comparison. The lower panels show the relative BSD difference between realisations with $i=10^8$ and $i=10^9$, respectively. Left and right panels demonstrate the situation of EOS Bright galaxies and Faint galaxies model, respectively.}
\label{figure: test_BSD_iterationnumberz9} 
\end{figure*}

The resolution of lensing and EoR simulations determines the lower limits of the BSD measurement with MFP, i.e., the parameter of the shortest travel distance of testing particles in MFP should be larger than the coarsest pixel size adopted in the above simulations because it reveals no physical information when travel distances of testing particles are smaller than pixel sizes \citep{2011MNRAS.411..955M}. Specifically, if we set the shortest travel distance to be 0.5 cMpc at $z_s = 14$ (HIGH-RES EoR simulation), approximately $40\%$ of the shortest travel distances of random testing particles in the MFP algorithm are less than this; in the EOS Bright Galaxies model, this fraction increases to about $65\%$, where the corresponding resolution is about 1.56 cMpc. When we measure BSDs of lensed EoR maps with the resolution of lensing simulations, the corresponding resolution is about 0.25 cMpc at $z_s = 14$. Conservatively, we choose 2 cMpc as the shortest travel distances in MFP, which is much smaller than the typical sizes of the largest bubbles (15 $\sim$ 25 cMpc at $z_s = 12$ and 10 $\sim$ 20 cMpc at $z_s = 14$). This setup does not affect our conclusion that lensing significantly affects the largest bubbles. However, to accurately investigate the lensing effects on small bubbles, we have to employ simulations of lensing and EoR with the required resolutions.

Besides, the measurement of BSDs via the MFP method involves intrinsic Monte Carlo sampling uncertainties, which means that the iteration count $i$ should be large enough to eliminate the influence due to shot noise. We conducted a test comparing BSD measurements across different iteration numbers to quantify potential numerical noise due to finite sampling. At $z_s>12$ and $z_s=9$, our baseline uses $i=10^6$ and $i=10^8$ Monte Carlo iterations, respectively; for validation we increase to $i=10^8$ (for $z_s>12$) and $i=10^9$ (for $z_s=9$). At the latter stage of EoR, the size of the ionised region is large and a higher iteration count $i$ is needed. We apply these tests to the HIGH-RES runs at $z_s=12$ and $14$, and to the EOS Bright and Faint galaxies models at $z_s=9$. We test both lensed and unlensed BSD. For $z_s=12$ and $14$, we recompute the lensed BSD for all 50 lens cones with $i=10^8$ (i. e., we totally reproduce the final result of HIGH-RES shown in Figure \ref{figure: bubble_size_pdf} using $i = 10^8$). At $z_s=9$, the $i=10^9$ test is only performed for randomly 5 of the 50 cones (i. e., 10\%) due to the large calculation resource requirement of the setting of $i = 10^9$. Figure \ref{figure: test_BSD_iterationnumber} and Figure \ref{figure: test_BSD_iterationnumberz9} show that the BSDs measured in the standard case and the validation case are very close, especially at $z_s$ = 12 and 14. Relative difference is $< 10\%$ except for the largest size end. Confirming that our choice of the number of iterations is reliable at these redshifts.

\section{Conclusions}\label{section: conclusions}

In this study, we conducted a suite of lensing ray-tracing simulations in the case of the multi-lens-plane with various ionisation models to investigate the impacts of lensing by galaxy clusters on ionisation structures during the EoR. By comparing the BSD of lensed and unlensed 2D \(x_{\mathrm{HI}}\) maps using the MFP method, we found that gravitational lensing significantly increases the number of bubbles at the large size end due to the lensing magnification. In contrast, the statistics of small-scale bubbles remain unaffected by lensing down to the bubble size of 2 cMpc throughout the entire EoR. Besides, different ionisation models yield varying results. Specifically, in the ionisation model dominated by the EOS Faint Galaxies model, the number of bubbles larger than 15 cMpc increases by 219\% at a redshift of 14. Conversely, in the model dominated by the EOS Bright Galaxies model, this number increases by 832\%.

In addition, the lensing effects on BSD change with redshift, as the influence of lensing magnification is sensitive to the topology of the ionisation bubbles. For instance, according to the redshift ranges defined in \cite{2024arXiv241108943J} which traces the ionised bubble merging path using data products from the THESAN simulations \citep{2022MNRAS.511.4005K} as sources, the evolution of ionised bubbles in a faint galaxies dominated reionisation model can be divided into three distinct stages: (\textrm{i}) Initial Expansion Phase ($z$ $\gtrsim$ 11); (\textrm{ii}) Merging Phase ($z$ $\approx$ 9-11); (\textrm{iii})  Rapid Expansion Phase ($z$ $\lesssim$ 9). For the case of the faint galaxy dominated models as sources in our lensing simulations, we found that the lensing effects on BSD significantly exist in the bubble Initial Expansion Phase and are suppressed in the Merging Phase and Rapid Expansion Phase, as illustrated in Figure \ref{figure: bubble_size_pdf} in Sect. \ref{section: results}.

In addition, further tests reveal the anisotropy of the lensing-induced effect, namely that the lensing effect is significant in the transverse-projected directions, but its impact on the LOS-projected MFP ray-segment length distribution is much weaker. This suggests that the LOS-projected MFP statistic may provide a useful route to recovering unlensed bubble statistics and deserves further study.

To verify the reliability of our conclusions, we examined how various simulation settings influence the outcomes. We quantified the lensing effects on the BSD using five ionisation simulations that varied in the minimum ionising source mass, ionisation efficiency, or ionisation simulation resolution (as illustrated in Sect. \ref{section: Reionization simulation generation}). The results across these simulations consistently demonstrated that lensing significantly impacts the BSD during the early stages of the EoR, specifically by increasing the expected number of large-scale bubbles. Moreover, we assessed the influence of different lensing simulation ray-tracing resolutions and iteration numbers in the MFP method used for BSD measurement, confirming the tendency of our findings, which are shown in Sect. \ref{section: uncertainties caused by lens light cone and ray tracing} and Sect. \ref{section: uncertainties caused by BSD measuring}. However, the pixelization of the mock source planes might underestimate the lensing effects on small-scale ionisation bubbles, which can be enhanced with future finer ionisation simulations. The box size of the mock might lead to statistical effects in the size distribution of large bubbles, as simultaneously achieving both high angular resolution and a large box size remains a challenge.

Notably, we adopt relatively simple assumptions regarding the lens and source populations to illustrate the tendency of the lensing effects on BSD by galaxy clusters. For instance, the source models contain different ionisation parameters, such as $M_{\rm min}$ and $\zeta$, leading to various lensing effects at given redshifts. Besides, the angular positions of the deflectors in our simulations are generated via the Monte Carlo method based on the assumption of uniform distribution without accounting for large-scale structures like cosmic filaments, which can be improved by employing haloes or even particles from N-body simulations directly. The mass distribution of the deflectors adheres to the TNFW model but does not include contributions from subhaloes, which can be addressed using Semi-analytic models or hydrodynamic simulations. Last but not least, while we present a theoretical exploration of lensing effects on BSD, it is critical to consider instrumental effects if one intends to constrain the statistics of sources and lenses with the lensing effects on BSD.

In summary, gravitational lensing caused by galaxy clusters leads to a non-negligible influence on the statistical properties of the ionisation structures during EoR, such as the size distribution of ionisation bubbles. The emergence of next-generation radio telescopes, such as the SKA, offering sufficient sensitivity and extensive sky coverage, will enable the observation of ionisation bubble statistics during the EoR as expected, which will promote astrophysical and cosmological investigations. Therefore, it is essential to carefully consider the lensing effects in relevant studies to ensure accurate and unbiased interpretations of observations.

\section*{Acknowledgements}

D.W. and N.L. acknowledge the support of the CAS Project for Young Scientists in Basic Research (No. YSBR-062). We thank Bin Liu and Yidong Xu for insightful discussions. HYS acknowledges the support from NSFC of China under grant 12533008, Ministry of Science and Technology of China (grant Nos. 2020SKA0110100), Key Research Program of Frontier Sciences, CAS, Grant No. ZDBS-LY-7013 and Program of Shanghai Academic/Technology Research Leader. Z.Z. acknowledges the support from the National Science Foundation of China (Grant No. 12203085). We thank ChatGPT for polishing the English in this paper.

\section*{Data Availability}

The data utilized in this study will be provided upon reasonable request.



\bibliographystyle{mnras}
\bibliography{refs}





\bsp	
\label{lastpage}
\end{document}